\documentclass[preprint,3p,12pt]{elsarticle}
\usepackage{mathrsfs}
\usepackage{amsmath}
\usepackage{stmaryrd}
\usepackage{bbding}
\usepackage{dcolumn}
\usepackage{graphicx}
\usepackage{amsfonts}
\usepackage{amssymb}
\usepackage{psfrag}
\usepackage{wrapfig}
\usepackage{subfigure}
\usepackage{makeidx}
\usepackage{bm}
\usepackage{epsf}
\usepackage{epsfig}
\usepackage{setspace}
\usepackage{graphicx}
\usepackage{epstopdf}
\usepackage{psfrag}
\usepackage{subfigure}
\usepackage{color}

\begin{document}
\title{Direct Modeling for Computational Fluid Dynamics \\
and the Construction of High-order Compact Scheme \\
for Compressible Flow Simulations}

\author[HKUST1]{Fengxiang Zhao}
\ead{fzhaoac@connect.ust.hk}

\author[HKUST2]{Xing Ji}
\ead{xjiad@connect.ust.hk}

\author[HKUST1]{Wei Shyy}
\ead{weishyy@ust.hk}

\author[HKUST1,HKUST2,HKUST3,HKUST4]{Kun Xu\corref{cor}}
\ead{makxu@ust.hk}

\address[HKUST1]{Department of Mechanical and Aerospace Engineering, Hong Kong University of Science and Technology, Clear Water Bay, Kowloon, HongKong}
\address[HKUST2]{Department of Mathematics, Hong Kong University of Science and Technology, Clear Water Bay, Kowloon, HongKong}
\address[HKUST3]{Guangdong-Hong Kong-Macao Joint Laboratory for Data-Driven Fluid Mechanics and Engineering Applications, Hong Kong University of Science and Technology, Hong Kong, China}
\address[HKUST4]{Shenzhen Research Institute, Hong Kong University of Science and Technology, Shenzhen, China}
\cortext[cor]{Corresponding author}

\begin{abstract}
Computational fluid dynamics is a direct modeling of physical laws in a discretized space.
The basic physical laws include the mass, momentum and energy conservations, physically consistent transport process, and similar domain of dependence and influence between the physical reality and the numerical representation.
Therefore, a physically soundable numerical scheme must be a compact one which involves the closest neighboring cells within the domain of dependence for the solution update under a CFL number $(\sim 1 )$.
In the construction of explicit high-order compact scheme,
subcell flow distributions or the equivalent degree of freedoms beyond the cell averaged flow variables must be evolved and updated, such as the gradients of the flow variables inside each control volume.
Under such a requirement, the direct use of Riemann solver as the evolution model is not adequate in its dynamics for the construction of high-order compact scheme. High-order dynamic process has to be modeled on the scales of cell size and time step.
The direct modeling of flow evolution under generalized initial condition will be developed in this paper.
In order to provide reliable cell averaged flow variables and their gradients for the compact data reconstruction,
the evolution process has to be able to provide discontinuous time-dependent flow variables across a cell interface.
At the same time, the time accurate flux function at a cell interface can become a discontinuous function of time.
Same as the spatial limiter in the conventional CFD methods, such as TVD and WENO, the temporal limiter for the flux function in time has to be
designed properly as well.
The direct modeling in this paper will provide the updates of flow variables differently on both sides of a cell interface and limit high-order time derivatives of the flux function nonlinearly in case of discontinuity in time, such as a shock wave moving across a cell interface within a time step.
The direct modeling unifies the nonlinear limiters in both space for the data reconstruction and time for the time-dependent flux transport.
The equivalent treatment of space and time makes the scheme be super robust and accurate for the compressible flow simulation.
At the same time, the high-order compact scheme can use a large CFL number $(\sim 0.8)$ in flow computation, even in the hypersonic flow simulation.
Under the direct modeling framework, as an example, the high-order compact gas-kinetic scheme (GKS) will be constructed.
The scheme shows significant improvement in terms of robustness, accuracy, and efficiency in comparison with the previous high-order compact GKS.

\end{abstract}


\maketitle

\section{Introduction}

The development of high-order compact schemes has attracted great attention in the past decades. Significant progress has been observed with the appearance of a wide variety of high-order schemes, such as weighted essentially non-oscillatory (WENO), discontinuous Galerkin (DG),
correction procedure via reconstruction (CPR), dispersion-relation preserving (DRP), high-order weighted compact nonlinear schemes (WCNS) etc. \cite{liu-WENO,jiang-WENO,reed,cockburn1,FR,CPR_wang,lele,DPR,WCNS}. However, the determinacy of the underlying principle and methodology for developing high-order compact scheme is still vague. Difficulties and inferior performance are often encountered in almost all existing high-order compact schemes, especially in the complicated flow simulations with shock interactions.
The design of delicate nonlinear limiters and trouble cell detection become the routine work in the pursuit of high-order compact schemes.
Here we will set up a framework for the construction of high-order compact scheme and provide necessary dynamic processes in the gas evolution model. The direct modeling includes the capturing of a discontinuous shock wave passing through a cell interface within a time step (never happens in the Riemann solver), the updating of multiple cell interface values,
and the limiting of  time derivatives of a discontinuous flux function in time (equivalent treatment in spatial reconstruction and  time evolution).
The numerical procedures of reconstruction, evolution, and projection in a second-order scheme will be transformed to the steps in the development of high-order scheme, and the emphasis is on the consistency of the evolution model and the solution update.

In this paper, Section 2 will present the basic principles of direct modeling in the construction of high-order compact schemes.
In Section 3, as an example, the high-order compact gas-kinetic scheme will be presented by following the principles.
Section 4 is the numerical examples which are used to validate the scheme.
The last section is the conclusion.

\section{Direct modeling for high-order compact scheme}

The fluid dynamic equations represent the physical laws of flow evolution in continuous space and time.
The domain of dependence and influence are determined by the wave propagating speed.
The computational fluid dynamics is about the discrete representation of physical conservation laws,
\begin{equation}\label{conservation}
\int_{\Omega_j} {\bf W} ({\bf x}, t^{n+1} ) \mathrm{d}V = \int_{\Omega_j} {\bf W} ({\bf x}, t^{n} ) \mathrm{d}V -
 \int_{t^n}^{t^{n+1}} \int_{\partial \Omega_j} {\bf F}(t) \cdot {\bf n} \mathrm{d}S,
\end{equation}
where $\bf W$ is the conservative flow variables, such as mass, momentum, and energy distributed in a control volume $ \Omega_j$ at discrete time
steps $t^n$ and $t^{n+1}$, and ${\bf F}(t)$ is the corresponding flux across the cell interface $\partial \Omega_j$.
The above conservation laws are valid in any space and time scales and in any flow regimes from the rarefied to the continuum one
once the dynamics of ${\bf F}(t)$ is properly modeled.
The quality of the scheme depends critically on the modeling of the time-dependent interface flux function ${\bf F}(t) $, which subsequently requires the reconstructed initial condition from ${\bf W}(t^n) $ and the evolution models of ${\bf W} (t)$ and
${\bf F}(t)$ at a cell interface.
For example,  the evolution models can be the Boltzmann solution in the rarefied flow regime and the Navier-Stokes one in the continuum flow regime \cite{xu2021}.
Due to the particle composition of fluid system, the physical propagation speed is coming from particle transport and collision,
which  has a limited value, such as the sound speed.
The direct modeling of ${\bf W} (t)$ and ${\bf F} (t)$ in Eq.(\ref{conservation})
should have the numerical domains of dependence and influence as close as possible to the physical one,
which is equivalent to determining time step $\Delta t = t^{n+1} - t^n$ with a CFL number on the order $(\sim 1)$, and including the closest neighboring cells in the reconstruction and evolution.
Therefore, only the compact schemes with neighboring interaction are the physically consistent numerical algorithms.

For a high-order compact scheme, the update of cell averaged flow variables alone in Eq.(\ref{conservation}) is not enough to
fully determine the local flow structure with high-order accuracy, such as on the order $(\geq 3 )$.
In order to recover a high-order flow distribution, large stencils are usually used in the reconstruction which contradicts with the
compactness requirement and the consistency between the physical and numerical domain of dependence.
Therefore, besides the updates of cell averaged flow variables, other variable related to the subcell resolution has to be evolved as well.
The strategy of using coupled flow variable and their derivatives to get high-order solution at an instant of time, such as the scheme of Lele,
is purely based on the numerical consideration, which is not the dynamic one.
The DG scheme uses the weak formulation to get the additional flow variables, such as the node values or derivatives of subcell flow variable, which is
more or less a dynamic model. The question in DG is about the reliability of the updated solution through the weak form.
As a direct modeling, here we are trying to use the "strong" solution or truly dynamic evolution solution to update the cell averaged gradients  inside each control volume, which can be equivalently considered as node values inside the control volume $\Omega_j$ as well.
Similar to Eq.(\ref{conservation}), the cell-averaged gradients can be updated based on the Gauss-Green theorem,
\begin{equation}\label{slope}
 \int _{\Omega_j} \nabla {\bf W} ({\bf x}, t^{n+1} ) \mathrm{d}V = {\Omega_j} (\nabla {\overline{\bf W}})_j =
 \int_{\partial \Omega_j} {\bf W}(t^{n+1}) {\bf n} \mathrm{d} S,
\end{equation}
where the flow variables $\bf W$ should be provided at the inner sides of the cell boundary of the control volume at the time step $t^{n+1}$.
Eq.(\ref{slope}) is mathematically precise under the assumption of smooth flow distribution inside each control volume $\Omega_j$.
In the discrete space, with the cell size resolution it is impossible to identify the subcell discontinuity except using the shock fitting.
As a shock capturing scheme, the appropriate approach is to assume a continuous subcell flow distribution and contribute
the possible discontinuity at the cell interface. Therefore, ${\bf W}_{j+1/2}$ may have multiple values, such as ${\bf W}_{j+1/2,-} $ and ${\bf W}_{j+1/2,+} $ at both sides of a cell interface under a single flux function ${\bf F}_{j+1/2} = {\bf F} ({\bf W}_{j+1/2,-}) = {\bf F} ({\bf W}_{j+1/2,+}) $. The outstanding example is that a shock is exactly located on the cell interface.
Therefore, in general the evolution model of the scheme should be able to update the values ${\bf W}_{j+1/2, \pm} (t^{n+1})$ separately in case of discontinuity, which requires a dynamic modeling rather than mathematical manipulation.
The updates of the flow variables and their gradients depend on the direct modeling of flux function and flow variables at the cell interface.
Their evolution model plays a dominant role for the quality of the scheme.
In general, the  flux function must be a time dependent function ${\bf F}(t)$ at a cell interface, and this function is not necessary to be a continuous function of time $t$ as analyzed later.

Based on the time-independent flux function $\bf F$ of the Riemann solver, the foundation of the second-order scheme has been established in 1970s and 1980s, where the nonlinear limiter in reconstruction of the initial data is the most important concept and practice for its success \cite{toro}.
In the traditional second order schemes, the basic numerical procedures are composed of the reconstruction, evolution, and projection.
The high-order compact scheme proposed in this paper is based on the governing equations (\ref{conservation}) and (\ref{slope}).
The direct modeling refers to the construction of the time evolution solution ${\bf W}(t)$ and ${\bf F}(t) $ in order to close these two equations in the updates of cell averaged flow variables and their gradients.
The direct modeling scheme can be also analyzed under the framework of reconstruction, evolution, and projection procedures.

\noindent{\bf Reconstruction}:

In the reconstruction stage, the most distinguishable achievement is the use of nonlinear limiters on either flux or flow variables.
For a second-order scheme, the reconstructed flow variables inside each control volume avoid the creation of local extreme and have the property of total variation of diminishing (TVD) \cite{harten1}.
Even for the linear equation, the scheme has to be designed nonlinearly.
Based on the reconstructed initial condition, theoretically a generalized Riemann problem (GRP) for the NS equations should be used as the dynamical model for the evaluation of time-dependent interface flux function in Eq.(\ref{conservation}).
However, due to its complexity in GRP, a Riemann solver for the inviscid part and a central difference for the viscous part are usually adopted in the construction of the flux function.  In order to get second order accuracy in time, the Runge-Kutta or Hancock method can be usually used in the evolution.
With the Riemann-solver based flux function, the cell averaged flow variables can be updated through the numerical conservation laws in Eq.(\ref{conservation}) for a second-order scheme.

For a high-order compact scheme, the stencil used should theoretically be the same as the second-order one.
Suppose that the compact scheme has reliable flow variables and their derivatives updated in Eq.(\ref{conservation}) and (\ref{slope}),
a reconstruction with compact stencil can be conducted through WENO or HWENO formulation. For example,
a 6th or 8th-order polynomial on structured mesh and a fourth-order one on unstructured mesh can be obtained \cite{CGKSAIA,zhao2020acoustic,compact-GKS-tri}.

\begin{figure}[!htb]
\centering
\includegraphics[width=0.6\textwidth]{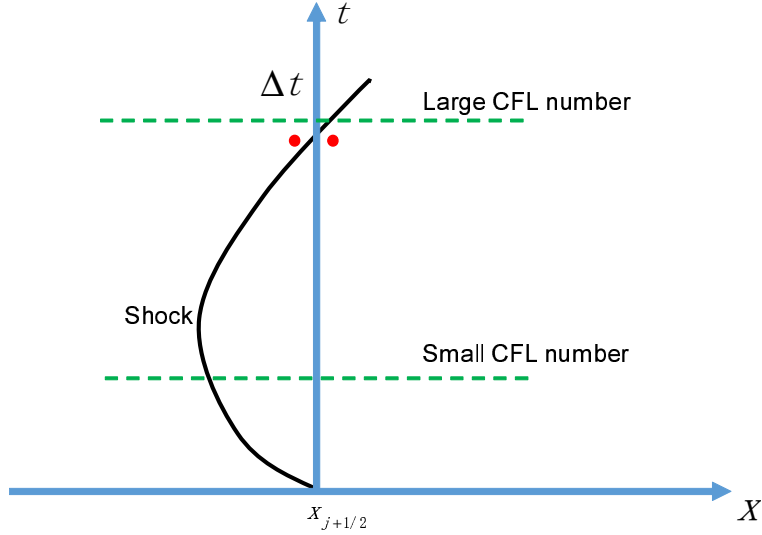}
\caption{\label{evolution1} A possible solution in the dynamical evolution process. At the dot points, there may have discontinuous ${\bf W}$ in space and discontinuous ${\bf F}$ in time.}
\end{figure}

\noindent{\bf Evolution}:

The evolution stage is the most critical one in the determination of the physical solution of ${\bf W}(t)$
and the flux function ${\bf F} (t) $ at a cell interface in order to close the system of Eq.(\ref{conservation}) and (\ref{slope}).
Starting from the reconstructed piecewise polynomial as the initial condition, a time accurate evolution solution should be modeled
in a discretized space.
Different from the Riemann solution, here the solution can be much more complicated in order to recover the physics, such as the NS solution and
the dynamic effect of a discontinuous shock. Certainly, the generalized Riemann solver helps, but it is limited to the
inviscid Euler solutions.
Here we require a time accurate flow variable and flux function, where the possible discontinuities in both flow variables and flux functions
in space and time have to be taken into account.
The time marching methods involving the
extra time derivatives can be used to achieve high-order temporal accuracy.
In order to fully utilize the time accurate evolution solution, the time-dependent flux function and its time derivative at different
intermediate  stages can be adopted to achieve high-order temporal accuracy.
Let $${\cal L}_j ({\bf W}) = -\frac{1}{|\Omega_j|}\int_{\partial \Omega_j} {\bf F}\cdot {\bf n} \mathrm{d}S ,$$ and
 $${\cal L}_{j,t} ({\bf W}) =-\frac{1}{|\Omega_j|} \int_{\partial \Omega_j} \frac{\partial}{\partial t}{\bf F}\cdot {\bf n} \mathrm{d}S ,$$
where the time-dependent flux function ${\bf F}(t)$ has to be modeled according to the flow regimes. In this paper, we are mainly
targeting on the NS solution in the continuum flow regime.
A high-order polynomial approximation of ${\cal L}_j$ in time can be obtained based on the evolution solutions at two stages $t=t_n$ and $t=t_n+\Delta t/2$.
The optimal approximation of the time integration of ${\cal L}_j$ is
\begin{align}\label{evolution-L}
\int_{t^n}^{t^{n+1}} {\cal L}_j (t) \mathrm{d}t=\Delta t {\cal L}_j ({\bf W}^n) +\frac{\Delta t^2}{2} {\cal L}_{j,t} ({\bf W}^n) -\frac{\Delta t^2}{3} {\cal L}_{j,t} ({\bf W}^n) +\frac{\Delta t^2}{3} {\cal L}_{j,t} ({\bf W}^{n+1/2}),
\end{align}
where the first two terms of the RHS corresponds to a second-order solution from the evolution at time stage $t^n$, and the last two terms are the high-order terms combing the evolution at middle stage $t^{n+1/2}$.
In addition, to get the evolution solution at $t^{n+1/2}$, the cell-averaged value and its derivative are required and updated by the evolution solution at $t^n$ based on Eq.(\ref{conservation}) and (\ref{slope}).
As a result, the update of ${\bf W}_{j}^{n+1/2}$ and ${\bf W}_{j}^{n+1}$ can be given as
\begin{align}\label{evolution}
\begin{split}
{\bf W}_{j}^{n+1/2} &= {\bf W}_j^n + \frac{1}{2} \Delta t {\cal L}_j ({\bf W}^n) + \frac{1}{8} \Delta t^2 {\cal L}_{j,t} ({\bf W}^n) , \\
{\bf W}_{j}^{n+1} &= {\bf W}_j^n + \Delta t {\cal L}_j ({\bf W}^n) + \frac{1}{6} \Delta t^2 ( {\cal L}_{j,t} ({\bf W}^n)
+ 2 {\cal L}_{j,t} ({\bf W}^{n+1/2})).
\end{split}
\end{align}
The detailed derivation will be given in Section 3. The above formula is the same as the standard two-stage fourth-order (S2O4) method for
the time discretization of semi-discrete conservation laws \cite{li}.
In the above temporal evolution model, a flux function depending continuously on time is assumed.
Physically, the flow evolution can be much more complicated than that, especially under a generalized high-order initial flow distribution.
Here there is the possibility in the flux function to appear discontinuity in time within a time step $\Delta t$, such as the
high order evolution model in Fig.(\ref{evolution1}), where
within the time step a discontinuous $\bf W$ in space and a discontinuous $\bf F$ in time may appear at the cell interface once a shock wave impinging on the cell interface.
The high-order evolution model as shown in Fig.(\ref{evolution1}) can appear in the S2O4 time discretization.
Therefore,  a reliable evolution model should be able to update two states, such as ${\bf W}^l ({\bf x}_{j+1/2}, \Delta t)$ and ${\bf W}^r ({\bf x}_{j+1/2}, \Delta t)$ at the left and right sides of the cell interface ${\bf x}_{j+1/2}$, for the update of the numerical solution in Eq.(\ref{slope}). A single value of ${\bf W} (t^{n+1}) $ at a cell interface is only a special case with the assumption of continuous flow evolution, which has been used before in the compact GKS \cite{ji2,CGKSAIA,ji2020-AIAA}.
In this paper, a generalized cell interface value will be evolved, which greatly improve the reliability of the
updated slope in Eq.(\ref{slope}).
 On the other hand, as a shock moving across the cell interface as shown in Fig.(\ref{evolution1}),
the flux function of ${\bf F}(t)$ at a cell interface is not a continuous function of time $t$ anymore and the above temporal evolution in Eq.(\ref{evolution-L}) or (\ref{evolution}) may become problematic for the update of the solution in Eq.(\ref{conservation}).
How to handle a possible discontinuous flux function
within a time step $[t^n , t^{n+1}]$ combined with the high-order time marching method is a problem we have
to solve in the construction of high-order schemes.

In order to find a proper way to handle the discontinuous flux function in time, similar to the slope limiter in space,
a nonlinear limiter in time is also needed. The limiting process of flux function in time is different from the nonlinear hybrid flux functions of
low order and high orders, such as the flux-corrected transport (FCT) method, or any other mixed fluxes in the implicit scheme \cite{FCT,DURAISAMY}.
The introduction of nonlinear time limiter in the flux function in this paper is to modify the high-order time derivatives in Eq.(\ref{evolution-L}) and the corresponding solution update in Eq.(\ref{evolution}).
This approach unifies the nonlinear limiting or reconstruction in both space and time which seem absent in the previous construction of high-order
 schemes, where great effort has been paid on the spatial reconstruction.
 More precisely, this practice extends the WENO-type concept in space to the temporal flux transport
 and the flow dynamics has been treated equivalently in the space and time for a shock wave propagating.
In other words, the above S2O4 method in Eq.(\ref{evolution}) for the temporal evolution has to be nonlinearly limited
in case of a discontinuous flux function in the high-order time evolution, where the assumption of
continuous flux function in time may be violated.
In this paper, a weighted limiter function will be introduced in Eq.(\ref{evolution}) in the high-order compact GKS in later sections.

Another commonly used time marching method is the multi-stage Runge-Kutta methods with the strong stability-preserving property (RK-SSP) which are developed for the time-independent numerical fluxes \cite{shu1988tvd,2001rk-ssp}.
The high-order accuracy and strong stability-preserving property are achieved by a convex combination of first-order forward Euler method in each stage.
However, such a technique cannot directly applied to the time-dependent flux function with high-order time derivatives, especially
with the possible discontinuity in time.
For a truly high-order compact scheme, similar to the space complexity the flux transport in time has the counterpart as well.
In addition, the strong stability-preserving (SSP) property can be hardly extended to evolution model with time derivatives in the flux function. Only limited progress has been reported for the multi-stage multi-derivative (MSMD) method with SSP property \cite{2016explicit-S2O4ssp}.

In the evolution stage,  the recipes introduced in this paper are the
limiting in the flux function ${\bf F}(t)$ in time and the constructing discontinuous
  solutions ${\bf W}(x_{j+1/2,\pm}$) at the cell interface in the updates of the solutions in Eq.(\ref{conservation}) and (\ref{slope}).
The development of a high-order compact GKS will be an example which is equipped with the above recipes.
As a result, the compact GKS has 4th-order accuracy in space and time in both structured and unstructured meshes,
shows super robustness in compressible flow simulation at high Mach number flow simulation, and takes a large CFL number, such as $0.8$ or above.

The stagnation in the development of high-order compact schemes in the CFD community is coming from the absence of high-order gas evolution model for the updating flow variables and their gradients in Eq.(\ref{conservation}) and (\ref{slope}). In other words, there are lack of reliable
evolution solutions ${\bf W}(t)$ and ${\bf F}(t)$ to close the modeling equations (\ref{conservation}) and (\ref{slope}), especially the
physical solution cannot be well-resolved by the numerical mesh size and time step.
The side effect for almost all existing high-order compact schemes, such as the lack of robustness and extremely limited CFL number, come from the inadequate dynamics in the evolution modeling, such as the simple adoption of the Riemann solver.
In this paper, we provide much more sophisticated dynamic model in the flow evolution.
For example, with the introduction of discontinuous cell interface values, the updated gradients become much more reliable for spatial reconstruction and make the scheme insensitive to the parameters in the reconstruction WENO methods.
With the use of nonlinearly time limited flux function, the high-order compact scheme can use a much large CFL number in the simulation, such as
$0.8$ in the fourth-order compact GKS instead of $0.1$ in the corresponding same order DG method.
Equipped with the above recipes in the direct modeling, the common practice in many high-order compact schemes, such as the trouble cell detection and additional delicate limiting processes, can be avoided.

\noindent{\bf Projection}:

With the above evolution model, the time accurate solutions of ${\bf W}(t)$ and ${\bf F}(t)$ at the cell interface can be used to
close the equations (\ref{conservation}) and (\ref{slope}).

According to the above reconstruction, evolution, and projection procedures,
the corresponding high-order compact GKS will be presented in the next section for the Euler and NS solutions.
The direct modeling in this section can be a good reference in the analysis of any other high-order compact scheme
and provide a vital approach for the further development of high-order compact schemes.

\section{High-order compact gas-kinetic scheme }

\subsection{Gas evolution model}
In this section, we are going to present the gas-kinetic scheme with all recipes introduced in the previous section, especially for the
updates of cell interface flow variables and the nonlinearly limited time-dependent flux function.
The gas-kinetic evolution model is based on the kinetic BGK equation \cite{BGK-1},
\begin{equation}\label{bgk}
f_t+\textbf{u}\cdot\nabla f=\frac{g-f}{\tau},
\end{equation}
where $\textbf{u}=(u,v)$ is the particle velocity, $f$ is the gas distribution function, $g$ is the corresponding equilibrium state that $f$ approaches, and $\tau$ is particle collision time.
The equilibrium state $g$ is a Maxwellian distribution,
\begin{equation*}
\begin{split}
g=\rho(\frac{\lambda}{\pi})^{\frac{K+2}{2}}e^{-\lambda((u-U)^2+(v-V)^2+\xi^2)},
\end{split}
\end{equation*}
where $\lambda =m/2kT $, and $m, k, T$ are the molecular mass, the Boltzmann constant, and temperature, respectively.
$K$ is the number of internal degrees of freedom, i.e. $K=(4-2\gamma)/(\gamma-1)$ for two-dimensional flow,
and $\gamma$ is the specific heat ratio. $\xi$ is the internal variable with $\xi^2=\xi^2_1+\xi^2_2+...+\xi^2_K$.
Due to the conservation of mass, momentum and energy during particle collisions, $f$ and $g$ satisfy the compatibility condition,
\begin{equation}\label{compatibility}
\int \frac{g-f}{\tau}\pmb{\psi} \mathrm{d}\Xi=0,
\end{equation}
at any point in space and time, where $\pmb{\psi}=(\psi_1,\psi_2,\psi_3,\psi_4)^T=(1,u,v,\displaystyle \frac{1}{2}(u^2+v^2+\xi^2))^T$, $\text{d}\Xi=\text{d}u\text{d}v\text{d}\xi_1...\text{d}\xi_{K}$.

The macroscopic mass $\rho$, momentum ($\rho U, \rho V$), and energy $\rho E$ can be evaluated from the gas distribution function,
\begin{equation}\label{g-to-convar}
{\textbf{W}} =
\left(
\begin{array}{c}
\rho\\
\rho U\\
\rho V\\
\rho E\\
\end{array}
\right)
=\int f \pmb{\psi} \mathrm{d}\Xi.
\end{equation}
The corresponding fluxes for mass, momentum, and energy in $i$-th direction is given by
\begin{equation}\label{g-to-flux}
{\textbf{F}_i} =\int u_i f \pmb{\psi} \mathrm{d}\Xi,
\end{equation}
with $u_1 = u$ and $u_2 = v$ in the 2D case.
On the mesh size scale, the conservation of mass, momentum and energy in a control volume has been given in Eq.(\ref{conservation}), and it is rewritten as
\begin{equation}\label{semifvs}
\textbf{W}^{n+1}_{j}=\textbf{W}^{n}_{j} +\int_{t^n}^{t^{n+1}} {\cal L}_j (t) \mathrm{d} t,
\end{equation}
where $\textbf{W}_{j}$ is the cell-averaged conservative variables defined as
\begin{align}\label{cell-average}
\textbf{W}_{j}&\equiv \frac{1}{\big| \Omega_j \big|} \iint_{\Omega_j} \textbf{W}(x,y) \text{d}x\text{d}y.
\end{align}
The line integral in ${\cal L}_j (t)$ is discretized by a q-point Gaussian integration formula,
\begin{align}\label{semifvs-rhs}
{\cal L}_j (t)=-\frac{1}{|\Omega_j|}\int_{\partial \Omega_j} \textbf{F}\cdot \textbf{n} \mathrm{d} S=-\frac{1}{|\Omega_j|} \sum_{l=1}^{l_0}\big( \sum _{k=1}^q \omega_k \textbf{F}(\mathbf{x}_k)\cdot \textbf{n}_l \big) \big|\Gamma_{l} \big|,
\end{align}
where $\textbf{F}=(\textbf{F}_1,\textbf{F}_2)$, $\big|\Gamma_{l} \big|$ is the side length of the cell, $l_0$ is the number of cell sides, $\textbf{n}_l$ is the unit outer normal vector, and $q$ and $\omega_k$ are the number of integration points and weight of the Gaussian integration formula.

In GKS, the evolution solution ${\bf W}(t)$ and ${\bf F}(t)$ at cell interface are determined by the time accurate gas distribution function $f$.
The integral solution  of BGK equation is \cite{xu2},
\begin{equation}\label{integral1}
\begin{split}
f(\textbf{x}_0,t,\textbf{u},\xi)=&\frac{1}{\tau}\int_0^t g(\textbf{x}',t',\textbf{u},\xi)e^{-(t-t')/\tau}\mathrm{d}t' \\
&+e^{-t/\tau}f_0(\textbf{x}_0-\textbf{u}(t-t_0),u,v,\xi),
\end{split}
\end{equation}
where $\textbf{x}_0$ is the numerical quadrature point at the cell interface, and $\textbf{x}_0=\textbf{x}^{'}+\textbf{u}(t-t^{'})$ is the particle trajectory.
Here $f_0$ is the initial state of gas distribution function $f$ at $t=0$.

In order to obtain the solution $f$, both $f_0$ and $g$ in Eq.(\ref{integral1}) need to be modeled.
Based on the integral solution, the gas distribution function with a second-order accuracy is \cite{xu2}
\begin{align}\label{2nd-f}
f(\textbf{x}_0,t,\textbf{u},\xi)=&(1-e^{-t/\tau})g_0+((t+\tau)e^{-t/\tau}-\tau)(\overline{a}_1u+\overline{a}_2v)g_0\nonumber\\
+&(t-\tau+\tau e^{-t/\tau}){\bar{A}} g_0\nonumber\\
+&e^{-t/\tau}g_r[1-(\tau+t)(a_{1r}u+a_{2r}v)-\tau A_r)]H(u)\nonumber\\
+&e^{-t/\tau}g_l[1-(\tau+t)(a_{1l}u+a_{2l}v)-\tau A_l)](1-H(u)),
\end{align}
where the terms related to $g_0$ are from the integral of the equilibrium state and the terms related to $g_l$ and $g_r$ are from the initial term $f_0$ in the Eq.(\ref{integral1}). All the coefficients in Eq.(\ref{2nd-f}) can be determined from the initially reconstructed  macroscopic flow variables at the left and right sides of the cell interface.
Higher-order evolution model, such as the third-order one, can be obtained as well \cite{li2010-HGKS}.

\subsection{Solution update at a cell interface}

In the previous compact GKS, based on the gas-kinetic evolution model in Eq.(\ref{2nd-f}) and S2O4 method for temporal discretization, both flow variables and the fluxes can be explicitly evaluated at each cell interface.
Then, the cell-averaged slope can be obtained with the flow variables at the cell interfaces by Green-Gauss theorem in Eq.(\ref{slope}).
In 1D case, flow variables variation in cell $j$ becomes
\begin{align}\label{Gauss-theorem}
{\bf W}_{j,x}^{n+1} \equiv \int_{I_j} \frac{\partial}{\partial x}{\bf W}(x,t^{n+1}) \mathrm{d}x=\frac{1}{\Delta x}({\bf W}_{j+1/2}^{n+1}-{\bf W}_{j-1/2}^{n+1}),
\end{align}
where ${\bf W}_{j+1/2}^{n+1/2}$ is the evolved solution at the cell interface.
Based on the both cell averaged flow variables and their slopes, the compact GKS with a single flow variable solution at a cell interface has been constructed \cite{CGKSAIA,compact-GKS-tri,ji2020-AIAA}.
One of the weakness in the above modeling is that a unique cell interface value ${\bf W}_{j+1/2}$ is assumed.
As analyzed in Section 2, even for a unique flux function $\bf F$, there may correspond to different $\bf W$, such as a shock front
is exactly located on the cell interface.
Since the initial condition for the flow variables are  piecewise discontinuous polynomials on both sides of the cell interface,
in the evolution process a discontinuous ${\bf W}_{j+1/2}$ may appear at the cell interface.
Based on the distribution function (\ref{2nd-f}), the flow variables at the cell interface can be obtained from the same
distribution function,
\begin{align}\label{point-evolution-e}
{\bf W}^e(x,t^{n+1})=\int f(\textbf{x}_0,t^{n+1},\textbf{u},\xi) \pmb{\psi} \mathrm{d}\Xi,
\end{align}
which are the macroscopic flow variables for the equilibrium state.
In the early compact GKS, the above ${\bf W}^e$ is used for the update of cell averaged slopes, where a smooth solution in space is assumed at
$({\bf x}_0, t^{n+1})$.
Under the non-equilibrium condition, discontinuities in the macroscopic flow variables may exist at a cell interface.
Based on the initial piecewise polynomials ${\bf W}^l(x,t^n), {\bf W}^r(x,t^n)$, two evolution solutions on both sides of the cell interface can be
constructed based on the assumption of independent evolution,
$$f^{l,r} ({\bf x}_0, t) = g^{l,r} ({\bf x}_0, t=0) ( 1 + A^{l,r} t ) ,$$
where the time evolution term $A^{l,r}$ is determined by the compatibility condition
$$\int (a^{l,r} u + A^{l,r}) \pmb{\psi} g^{l,r} \mathrm{d}\Xi = 0$$
and $a^{l,r}$ is given by
$$ \frac{\partial}{\partial x} {\bf W}^{l,r} = \int a^{l,r}  \pmb{\psi} g^{l,r} \mathrm{d} \Xi .$$
More detailed formulation can be found in \cite{xu1}.
 Based on the  independently evolved non-equilibrium gas distribution functions at the left and right sides of a cell interface, the corresponding macroscopic flow variables can be evaluated,
\begin{equation}\label{point-evolution-lr}
{\bf W}_0^{l,r}(x,t^{n+1})= \int \pmb{\psi} f^{l,r} \mathrm{d}\Xi . \\
\end{equation}
The final evolution solution of flow variables at cell interface are modeled as
\begin{align}\label{Gauss-theorem}
\begin{split}
{\bf W}^l(x,t^{n+1})&= (1-e^{-\Delta t/\tau_0}){\bf W}^e(x,t^{n+1}) +e^{-\Delta t/\tau_0} {\bf W}_0^l(x,t^{n+1}), \\
{\bf W}^r(x,t^{n+1})&= (1-e^{-\Delta t/\tau_0}){\bf W}^e(x,t^{n+1}) +e^{-\Delta t/\tau_0} {\bf W}_0^r(x,t^{n+1}).
\end{split}
\end{align}
The contributions from ${\bf W}_0^l$, ${\bf W}^r$, and ${\bf W}^e(x,t^{n+1})$ in the determination of ${\bf W}^l$ and ${\bf W}^r$
depend on the ratio of time step $\Delta t$ to the relaxation time $\tau$ for the local solution to approaching an equilibrium state.
 The weighting function $e^{-\Delta t/\tau}$ is consistent with the physical relaxation process used in GKS.
In the smooth flow region and in the continuum flow regime, $\Delta t \gg \tau$ holds.
Here, in the above equation, the relaxation time $\tau_0$ is defined as
\begin{align*}
\tau_0=C \displaystyle|\frac{p_l-p_r}{p_l+p_r}|\Delta t,
\end{align*}
where $C$ is a constant coefficient and a uniform  value $5$ is used in this paper.
In the smooth flow region, the initially reconstructed pressure $p_{l,r}$ at the left and right hand sides of a cell interface will be close to
each other and a single equilibrium macroscopic flow variables will be approached at the cell interface.

\subsection{Nonlinear limiter on temporal discretization}

The high-order temporal discretization can be developed through the multi-stage advancing process, such as four-stage fourth-order Runge-Kutta method (RK4) and two-stage fourth-order (S2O4) method  based on time-accurate numerical flux function \cite{li,liAIA}, as shown in Eq.(\ref{evolution}).
In the above time evolution process, the possible discontinuity of the flux function in time has not been considered.
In practice,  the flux function may become discontinuous in time, as shown in Fig.(\ref{evolution1}).
Without properly handling the discontinuity, the CFL number can be limited to a small value.
In order to increase the accuracy and robustness in the modeling of the flux function,
a nonlinear limiter for temporal discretization will be designed.
As an example, the S2O4 formulation will be modified. The idea can be extended to other high-order time evolution model as well.

In order to understand the limiting process, the high-order time discretization in Eq.(\ref{evolution})
will be derived in detail in a different way from the previous one \cite{li,pan1}.
To discretize the time integration of ${\cal L}_j (t)$ in Eq.(\ref{semifvs}) in a time step $[t^n,t^{n+1}]$, a high-order interpolation approximation of ${\cal L}_j (t)$ is defined first.
For simplicity and without loss of generality, a time step $[0,\Delta t]$ is considered in this section.
For a fourth-order time discretization, a third-order polynomial is constructed
\begin{equation}\label{p4_t}
\begin{split}
P^{3}(t)&=\sum_{k=0}^{3} b_k\frac{t^k}{k!} \\
        &=L_j(t)+O(\Delta t^4),
\end{split}
\end{equation}
where $L_j$ is one of the components of ${\cal L}_j$.
In order to determine the polynomial, the following conditions are given.
\begin{equation}\label{p4_t_bk1}
\begin{split}
P^{3}(0)&=b_0=L_j^{n}, \\
P^{3}_t(0)&=b_1=L^n_{j,t}, \\
\end{split}
\end{equation}
\begin{equation}\label{p4_t_bk2}
\begin{split}
P^{3}(\Delta t/2)&=\sum_{k=1}^{3} b_k\frac{(\Delta t/2)^{k-1}}{(k-1)!}=L^{n+1/2}_{j}, \\
P^{3}_{t}(\Delta t/2)&=\sum_{k=2}^{3} b_k\frac{(\Delta t/2)^{k-2}}{(k-2)!}=L^{n+1/2}_{j,t}. \\
\end{split}
\end{equation}
Thus $b_2$ and $b_3$ can be uniquely determined as
\begin{equation}\label{p4_bk}
\begin{split}
b_2&=\frac{4}{t^2}\big(-\Delta t L_t^{n+1/2}+6L^{n+1/2}-2\Delta t b_1-6b_0\big),\\
b_3&=\frac{24}{t^3}\big(\Delta t L_t^{n+1/2}-4L^{n+1/2}+\Delta t b_1+4b_0\big),
\end{split}
\end{equation}
and the integration of $L_j(t)$ is obtained as
\begin{equation}\label{S2O4-linear}
\begin{split}
\int_0^{\Delta t} L_j(t) \mathrm{d}t =& \int_0^{\Delta t} P^3(t) \mathrm{d}t +O(\Delta t^5) \\
         =&\Delta t L^n_{j} + \frac{\Delta t^2}{2} L^n_{j,t} \\
          &-\frac{\Delta t^2}{3}L^n_{j,t}+\frac{\Delta t^2}{3}L^{n+1/2}_{j,t} +O(\Delta t^5),
\end{split}
\end{equation}
where the third and fourth terms of the RHS in Eq.\eqref{S2O4-linear} are from the higher-order terms (third- and fourth-order terms) of $P^3(t)$.
Based on the conservation laws of Eq.(\ref{semifvs}), the update of $\mathbf{W}_j^{n+1}$ is given as
\begin{equation}\label{S2O4-linear-1}
\begin{split}
\mathbf{W}_j^{n+1} =&\mathbf{W}_j^n +\Delta t\mathcal{L}_{j}(\mathbf{W}^n) + \frac{\Delta t^2}{2}\mathcal{L}_{j,t}(\mathbf{W}^n) \\
           &-\frac{\Delta t^2}{3}\mathcal{L}_{j,t}(\mathbf{W}^n)+\frac{\Delta t^2}{3}\mathcal{L}_{j,t}(\mathbf{W}^{n+1/2}),
\end{split}
\end{equation}
$\mathcal{L}_{j,t}^{n+1/2}$ can be determined by the time-accurate solver based on the intermediate stage variable $\mathbf{W}_j^{n+1/2}$ which is updated by Eq.(\ref{semifvs}) in the time step $[0,\Delta t/2]$ as
\begin{equation*}
\mathbf{W}_j^{n+1/2} =\mathbf{W}_j^n +\int_0^{\Delta t/2} {\cal L}_j(t) \mathrm{d}t,
\end{equation*}
and the evolution solution can be determined as
\begin{equation*}
{\cal L}_j(t) =\mathcal{L}_{j}^n +t\mathcal{L}_{j,t}^n,
\end{equation*}
where $\mathcal{L}_{j}^n$ and $\mathcal{L}_{j,t}^n$ have been determined before.
Thus the update of $\mathbf{W}_j^{n+1/2}$ is
\begin{equation}\label{S2O4-linear-0}
\mathbf{W}_j^{n+1/2} =\mathbf{W}_j^n +\frac{\Delta t}{2}\mathcal{L}_{j}(\mathbf{W}^n) + \frac{\Delta t^2}{8}\mathcal{L}_{j,t}(\mathbf{W}^n).
\end{equation}
Eq.\eqref{S2O4-linear-0} gives a third-order approximation to $\mathbf{W}_j(\Delta t/2)$, which is
\begin{equation*}
\mathbf{W}_j^{n+1/2}=\mathbf{W}_j(\Delta t/2)+O(\Delta t^3).
\end{equation*}
Substituting $\mathbf{W}_j^{n+1/2}$ into $\mathcal{L}_{j,t}^{n+1/2}$ into Eq.\eqref{S2O4-linear-1}, the error of $\mathbf{W}_j^{n+1}$ maintains $O(\Delta t^5)$.
Therefore, Eq.\eqref{S2O4-linear-0} and Eq.\eqref{S2O4-linear-1} together give a temporal discretization with a fourth-order accuracy for the conservation laws Eq.\eqref{semifvs}, and the formulations are the same as those in Eq.(\ref{evolution}), the so-called S2O4 method  \cite{li,seal}.
The time derivative of numerical flux is required to obtain $\mathcal{L}_{j,t}$ and it is given by the time-accurate flux solver based on the initial piecewise polynomials, for details refer to \cite{pan1,compact-GKS-tri}.

The high-order approximation in Eq.\eqref{S2O4-linear-1} or S2O4 method is valid when $\mathbf{F}(t)$ and $\mathcal{L}_{j}(t)$
are continuous function of time. In the general case, as analyzed in Section 2, $\mathbf{F}(t)$ can be discontinuous in time,
and the high-order expansion in the temporal discretization in Eq.\eqref{S2O4-linear-1} may be flawed.
To obtain a robust high-order temporal discretization without reducing the CFL number, the nonlinear temporal discretization is developed.
The basic idea for designing nonlinear limiter in time is to switch the high-order evolution model to the second-order one
in case of discontinuous solution, where the high-order evolution model is obtained by unifying the evolutions at two time stages within a time step. In the basis for the above limiting process on the high-order time derivatives is that the one-step second-order evolution model
will not have the mechanism described in Fig.(\ref{evolution1}).
In order to limit the time evolution, the S2O4 method can be re-written as
\begin{equation}\label{S2O4-nonlinear}
\begin{split}
\mathbf{W}_j^{n+1} =&\mathbf{W}^n +\Delta t\mathcal{L}_{j}(\mathbf{W}^n) + \frac{\Delta t^2}{2}\mathcal{L}_{j,t}(\mathbf{W}^n) \\
           &-\frac{\Delta t^2}{3}\widetilde{\mathcal{L}}_{j,t}(\mathbf{W}^n)+\frac{\Delta t^2}{3}\widetilde{\mathcal{L}}_{j,t}(\mathbf{W}^{n+1/2}),
\end{split}
\end{equation}
where $\widetilde{\mathcal{L}}_{j,t}(\mathbf{W}^n)$ and $\widetilde{\mathcal{L}}_{j,t}(\mathbf{W}^{n/2})$ are the limited time derivatives.
Based on $\mathcal{L}_{j}$, $\widetilde{\mathcal{L}}_{j}$ is constructed as
\begin{equation}\label{semifvs-RHS-nonlinear}
\widetilde{\mathcal{L}}_{j}(\textbf{W}) = -\frac{1}{\big|\Omega_j\big|} \sum_{l=1}^{l_0} \omega^t_l \big( \sum _{k=1}^q \omega_k \textbf{F}(\mathbf{x}_k)\cdot \textbf{n}_l \big) \big|\Gamma_{l} \big|,
\end{equation}
where $\omega^t_l$ is a nonlinear weight for the $l$th side of the cell and $\omega^t_l \in [0,1]$.
Eq.\eqref{S2O4-linear-0} and Eq.\eqref{S2O4-nonlinear} together give a nonlinear S2O4 formula for the conservation laws of Eq.\eqref{semifvs}.
A small value of $\omega_l^t$ corresponds to a strong discontinuity of the flux function in time.
In smooth regions, $\omega_l^t$ is approximately equal to $1$ with high-order accuracy, and in the region near shock waves, $\omega_l^t$ tends to $0$.
In this paper, $\omega_l^t$ is defined as
\begin{equation}\label{S2O4-w}
\begin{split}
\widetilde{\alpha}^k_1&=1+ \big( \frac{\tau^k_Z}{IS^k_{min}+\epsilon}  \big)^2, ~~\widetilde{\alpha}^k_2=1+ \big( \frac{\tau^k_Z}{IS^k_{max}+\epsilon}  \big)^2, ~~k=L,R, \\
\alpha^k_2&=2\frac{\widetilde{\alpha}^k_2}{\widetilde{\alpha}^k_1+\widetilde{\alpha}^k_2}, \\
\omega_l^t&=\mathrm{min}\{\alpha^L_2,\alpha^R_2\},
\end{split}
\end{equation}
where $IS^{L,R}_{min}$ and $IS^{L,R}_{max}$ are the minimum and maximum smooth indicators in the nonlinear compact spatial reconstruction in the cells on both sides of the $l$th side, and $\tau^k_Z$ is the corresponding local higher-order reference value to indicate smoothness of the large stencil in the spatial reconstruction. $\epsilon$ is a small positive number which takes a value $1\times10^{-5}$ for all  numerical tests in this paper. The nonlinear compact spatial reconstruction will be given in the next section.
In smooth regions, the current nonlinear temporal discretization formula can return to the linear one in the sense of accuracy. By Taylor series expansion, there is
\begin{equation*}
\widetilde{\alpha}^k_i=C_0 \big(1+O(h^{2(r_{\tau}-2)+r})\big),
\end{equation*}
where $C_0$ is a constant parameter, $h$ is the cell size, $r_{\tau}$ is the order of $\tau^k_Z$, and $r$ is the order of the corresponding polynomial for obtaining $IS_{min}$ (with $i=1$) or $IS_{max}$ (with $i=2$). For example, if $IS_{min}$ is determined by a linear polynomial, then $r=1$. The minimum order of the candidate polynomials is $1$ in the nonlinear reconstruction adopted in this paper. $\tau^k_Z$ has at least the order of $O(h^3)$, and it will be defined in the next section. In smooth regions $\omega_l^t$ becomes
\begin{equation*}
\omega_l^t=1+O(h^3).
\end{equation*}
Thus, the truncation error introduced by $\omega_l^t$ in Eq.\eqref{S2O4-nonlinear} is $O(h^3\Delta t^2)$, and the nonlinear S2O4 formula has a fourth-order accuracy in smooth region.
In the region near shock wave, suppose the size of the discontinuities is always $O(1)$, and the order of $\omega_l^t$ becomes
\begin{equation*}
\omega_l^t=O(h^4).
\end{equation*}
Thus the high-order temporal interpolation for ${\bf W}_j(t)$ and $\mathcal{L}_{j}(t)$ in Eq.\eqref{S2O4-nonlinear} returns to the lower-order one.
For two-dimensional flows, four $\omega^t_l$ at each interface can be obtained corresponding to the four characteristic variables if the characteristics reconstruction is adopted. Since the shock wave can only correspond to the first or fourth characteristic variables, the unique $\omega^t_l$ of each interface can be obtained by taking the minimum one from the first and fourth variables.
The analysis and modification of the numerical scheme in this paper are well verified by numerical tests. The second-order temporal discretization is proved to have good robustness in numerical tests, thus only the higher-order terms (third order and fourth order) in the evolution process is modified by nonlinear weighting function in Eq.\eqref{S2O4-nonlinear}.
In a special case of a stationary shock wave at a cell interface, the flux function is a continuous function of time and the scheme works well even without the nonlinear limiting process.

\subsection{Compact reconstruction}

 The high-order compact spatial reconstruction will be briefly presented.
 Compared with the compact schemes without updating different cell interface flow variables and limiting time-dependent flux function \cite{CGKSAIA,compact-GKS-tri}, the reconstructions in the current new scheme become simpler and the solution will not be sensitive to the
 reconstruction schemes due to the reliable evolution model and accurate cell averaged slopes.
Several reconstruction schemes behave equally well.
In this section, a simple reconstruction scheme will be  given and used for all numerical tests.
Based on the candidate polynomials, the nonlinear combination is
\begin{equation}\label{C-ENO-WENO}
\begin{split}
R(\mathbf{x})=\sum_{k=1}^{n}w_k q_k(\mathbf{x}) + w_0 \big( \frac{1+C}{C}P(\mathbf{x}) -\sum_{k=1}^{n} \frac{C_k}{C}q_k(\mathbf{x}) \big),
\end{split}
\end{equation}
where $P(\mathbf{x})$ is the high-order polynomial obtained by the large stencil, and $q_k(\mathbf{x})$ are the lower-order polynomials obtained by the sub-stencils. The formula is a nonlinear combination of the high-order and lower-order polynomials, i.e., the so-called a combination of ENO and WENO methodology.
The nonlinear weights $w_k$ are
\begin{equation}\label{nonlinear-wk}
\begin{split}
&w_{k}=\frac{ \widetilde{w}_{k} }{ \sum_{j=0}^{n} \widetilde{w}_{j}  }, \\
&\widetilde{w}_{k}=\overline{d}_k \big( 1+ \big( \frac{\tau_Z}{IS_k+\epsilon}  \big)^2 \big),
\end{split}
\end{equation}
and $\overline{d}_{k}$ are
\begin{equation}\label{linear-dk}
\overline{d}_{0}=\frac{C}{1+C}, ~~ \overline{d}_k=\frac{C_k}{1+C},~~ k=1,\cdots,n,
\end{equation}
where $\epsilon$ is a small positive number with a value $1\times 10^{-5}$ for all numerical tests in this paper, $n$ is the number of the sub-stencils, and $\tau_Z$ is the local higher-order reference value to indicate smoothness of the large stencil given by $IS_k$ \cite{WENO-Z}.
Coefficients $C$ and $C_k$ are required to satisfy
\begin{equation*}
\sum_{k=1}^{k=n} C_k=1, ~~ C>0.
\end{equation*}
The compact scheme based on the current nonlinear reconstruction is insensitive to the values of $C$ and $C_k$. $C=5$ and $C_k=1/n$ used in \cite{compact-GKS-tri} can work very well.
In addition, the candidate polynomial $q_0$ is a higher-order one, and the calculation of $IS_0$ makes the high-order compact reconstruction less efficient. An efficient and robust technique is to take $IS_0=\mathrm{max}\{IS_1, IS_2, \cdots, IS_n\}$, which has been confirmed by the numerical examples in this paper.

\subsubsection{Compact reconstruction on structured mesh}

\begin{figure}[!htb]
\centering
\includegraphics[width=0.50\textwidth]{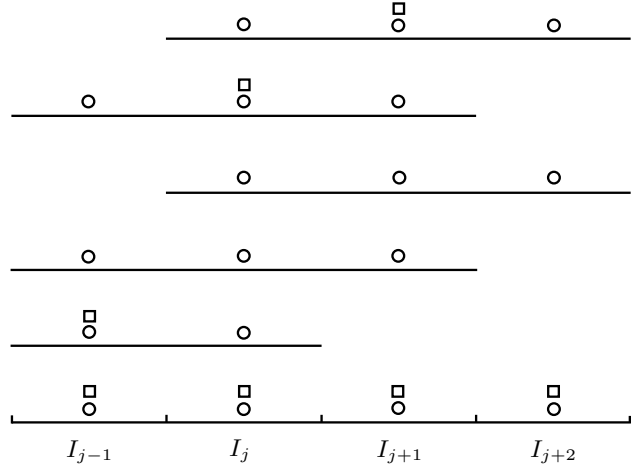}
\caption{\label{1d-stencil} A schematic of candidate stencils (including the large stencil and sub-stencils) for left side value at the cell interface $x_{j+1/2}$ in 8th-order compact GKS: The square represents the cell-averaged slope, and the circle represents the cell average of flow variables. One high-order large stencil and five low-order stencils are used.}
\end{figure}

The 8th-order compact GKS has been developed in \cite{CGKSAIA}. The high resolution and good robustness of the scheme for compressible flow and
aeroacoustic simulations have been validated \cite{CGKSAIA,zhao2020acoustic}.
Fig. \ref{1d-stencil} presents a schematic of the stencil selection for left side value at the cell interface $x_{j+1/2}$ in the current 8th-order compact reconstruction. Totally, one large stencil and five sub-stencils are used,
\begin{align*}
S_0&=\{Q_{j-1},Q_{j},Q_{j+1},Q_{j+2},Q^{'}_{j-1},Q^{'}_{j},Q^{'}_{j+1},Q^{'}_{j+2}\} \\
S_1&=\{Q_{j-1},Q_{j},Q^{'}_{j-1}\}, ~~S_2=\{Q_{j-1},Q_{j},Q_{j+1}\}, ~~S_3=\{Q_{j},Q_{j+1},Q_{j+2}\}, \\
S_4&=\{Q_{j-1},Q_{j},Q_{j+1},Q^{'}_{j}\}, ~~S_5=\{Q_{j},Q_{j+1},Q_{j+2},Q^{'}_{j+1}\}. \\
\end{align*}
All stencils appear in the previous 8th-order compact reconstruction and the detailed formulas of the reconstructed values at the cell interface and the smoothness indicators can be found in \cite{CGKSAIA}, where
 a 7th-order polynomial can be determined from $S_0$, three quadratic polynomials from $S_1$ $S_2$ and $S_3$, and two cubic polynomials from $S_4$ and $S_5$.
The nonlinear reconstruction is given in Eq.\eqref{C-ENO-WENO}. The local reference value $\tau_Z$ is constructed as \cite{CGKSAIA}
\begin{equation}\label{1d-recons-tau}
\tau_Z=\left|3(IS_{1}-IS_{2}) + (IS_{3}-IS_{2})\right|.
\end{equation}

\subsubsection{Compact reconstruction on triangular mesh}

\begin{figure}[!htb]
\centering
\includegraphics[width=0.45\textwidth]{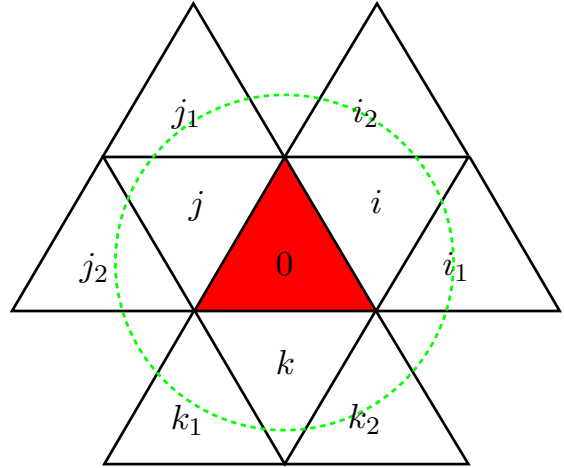}
\caption{\label{2d-stencil} A schematic of the large reconstruction stencil in compact GKS: The green dotted circle is a schematic of the largest physical domain of dependence at a reasonable CFL number, that is, the fluid element in cell $0$ may interact with the fluid element in the range of the circle. In each cell of the stencil, three datum, i.e., one cell averages and two cell-averaged derivatives, are known.}
\end{figure}

\begin{figure}[!htb]
\centering
\includegraphics[width=0.3\textwidth]{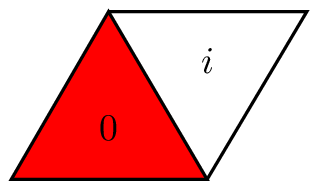}
\includegraphics[width=0.3\textwidth]{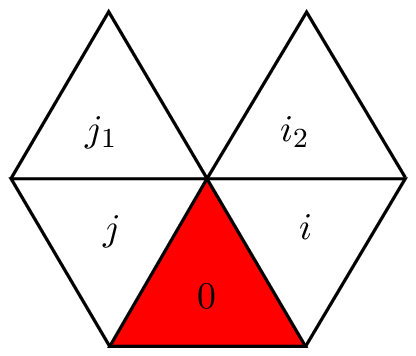}
\caption{\label{2d-stencil-1} A schematic of two of the sub-stencils in compact GKS: The left is a schematic of $S_1$, and the right is a schematic of $S_4$. Because the cell-averaged derivatives are known, a linear polynomial and a quadratic polynomial can be obtained by these two sub-stencils respectively based on least square method. }
\end{figure}

A class of 4th-order compact GKS on triangular mesh has been developed in \cite{compact-GKS-tri}.
Here a simple sub-stencil selection will be presented.
Fig. \ref{2d-stencil} shows a schematic of the large stencil in the compact GKS, where
the compactness means the compatibility between the physical and numerical domain of dependence.
The candidate stencils for the nonlinear reconstruction are given as follows.
\begin{align*}
S_0&=\{Q_0,Q_i,Q_j,Q_k,Q_{i_1},Q_{i_2},Q_{j_1},Q_{j_2},Q_{k_1},Q_{k_2},\nabla Q_{0},\nabla Q_{i},\nabla Q_{j},\nabla Q_{k}\}, \\
S_1&=\{Q_0,Q_i,\nabla Q_{i}\}, ~~~~~~S_2=\{Q_0,Q_j,\nabla Q_{j}\}, ~~~~~~S_3=\{Q_0,Q_k,\nabla Q_{k}\}, \\
S_4&=\{Q_0,Q_i,Q_j,Q_{i_2},Q_{j_1},\nabla Q_{0}\}, ~S_5=\{Q_0,Q_j,Q_k,Q_{j_2},Q_{k_1},\nabla Q_{0}\}. \\
S_6&=\{Q_0,Q_k,Q_i,Q_{k_2},Q_{i_1},\nabla Q_{0}\}, \\
\end{align*}
A cubic polynomial can be determined by $S_0$, three linear polynomials from  $S_1$ $S_2$ and $S_3$, and three quadratic polynomials
from $S_4$ $S_5$ and $S_6$. The least square method is used to determine these candidate polynomials and the details can be found in \cite{compact-GKS-tri}.
In the nonlinear reconstruction given by Eq.\eqref{C-ENO-WENO}, the local reference value $\tau_Z$ is constructed as
\begin{equation}\label{2d-recons-tau}
\tau_Z=\big(\left|2IS_0-IS_1-IS_2 \right| +\left|2IS_0-IS_2-IS_3 \right| +\left|2IS_0-IS_3-IS_1 \right|\big)/3.
\end{equation}

\section{Numerical examples}

Here we are going to validate the compact GKS on both structured and unstructured meshes.
The time step is determined by the CFL condition with $CFL=0.8$ in all test cases if not specified.
For viscous flow, the time step is also limited by the viscous term as $\Delta t=0.8h^2/(2\nu)$ as well, where $\nu$ is the kinematic viscosity coefficient and $h$ is the cell size. For one-dimensional uniform mesh, $h=\Delta x$; and for the two-dimensional triangular mesh,
$h_j=|\Omega_j |/2 |\Gamma_j|$ which is the radius of the inscribed circle in the control volume $\Omega_j$.
In all test cases, the same nonlinear reconstruction is used. There is no additional "trouble cell" detection and additional limiter designed for any specific test.

The collision time $\tau$ for the inviscid flow is defined by
\begin{align*}
\tau=\varepsilon \Delta t + C \displaystyle|\frac{p_l-p_r}{p_l+p_r}|\Delta t,
\end{align*}
where $\varepsilon=0.05$, $C=5$, and $p_l$ and $p_r$ are the pressure at the left and right sides of a cell interface.
For the viscous flow, the collision time is related to the viscosity coefficient,
\begin{align*}
\tau=\frac{\mu}{p} + C \displaystyle|\frac{p_l-p_r}{p_l+p_r}|\Delta t,
\end{align*}
where  $\mu$ is the dynamic viscosity coefficient and $p$ is the pressure at the cell interface.

\subsection{Accuracy test}
The one-dimensional advection of density perturbation is tested first. The initial condition is given as follows
\begin{align*}
\rho(x)=1+0.2\sin(\pi x),\ \  U(x)=1,\ \ \  p(x)=1, x\in[0,2].
\end{align*}
The periodic boundary condition is adopted. The results from the current 8th-order compact GKS are listed in
Table \ref{1d-accuracy-8-2}.

\begin{table}
	\begin{center}
		\def\temptablewidth{0.90\textwidth}
		{\rule{\temptablewidth}{0.70pt}}
		\begin{tabular*}{\temptablewidth}{@{\extracolsep{\fill}}c|cc|cc}
			
			mesh length & $L^1$ error & Order & $L^{\infty}$ error & Order  \\
			\hline
            1/10 & 2.7627e-04 &       & 5.6985e-04 &      \\
            1/20 & 3.5968e-06 & 6.26  & 8.1259e-06 & 6.13 \\
            1/40 & 1.2672e-08 & 8.15  & 6.3630e-08 & 7.00 \\
            1/80 & 5.9689e-11 & 7.73  & 1.5194e-10 & 8.71 \\
		\end{tabular*}
		{\rule{\temptablewidth}{0.1pt}}
	\end{center}
	\vspace{-1mm} \caption{\label{1d-accuracy-8-2} 1-D accuracy test: errors and convergence orders of the current 8th-order compact GKS with nonlinear reconstruction and $\Delta t = 0.5 \Delta x^2$.}
\end{table}

\begin{figure}[!htb]
\centering
\includegraphics[width=0.485\textwidth]{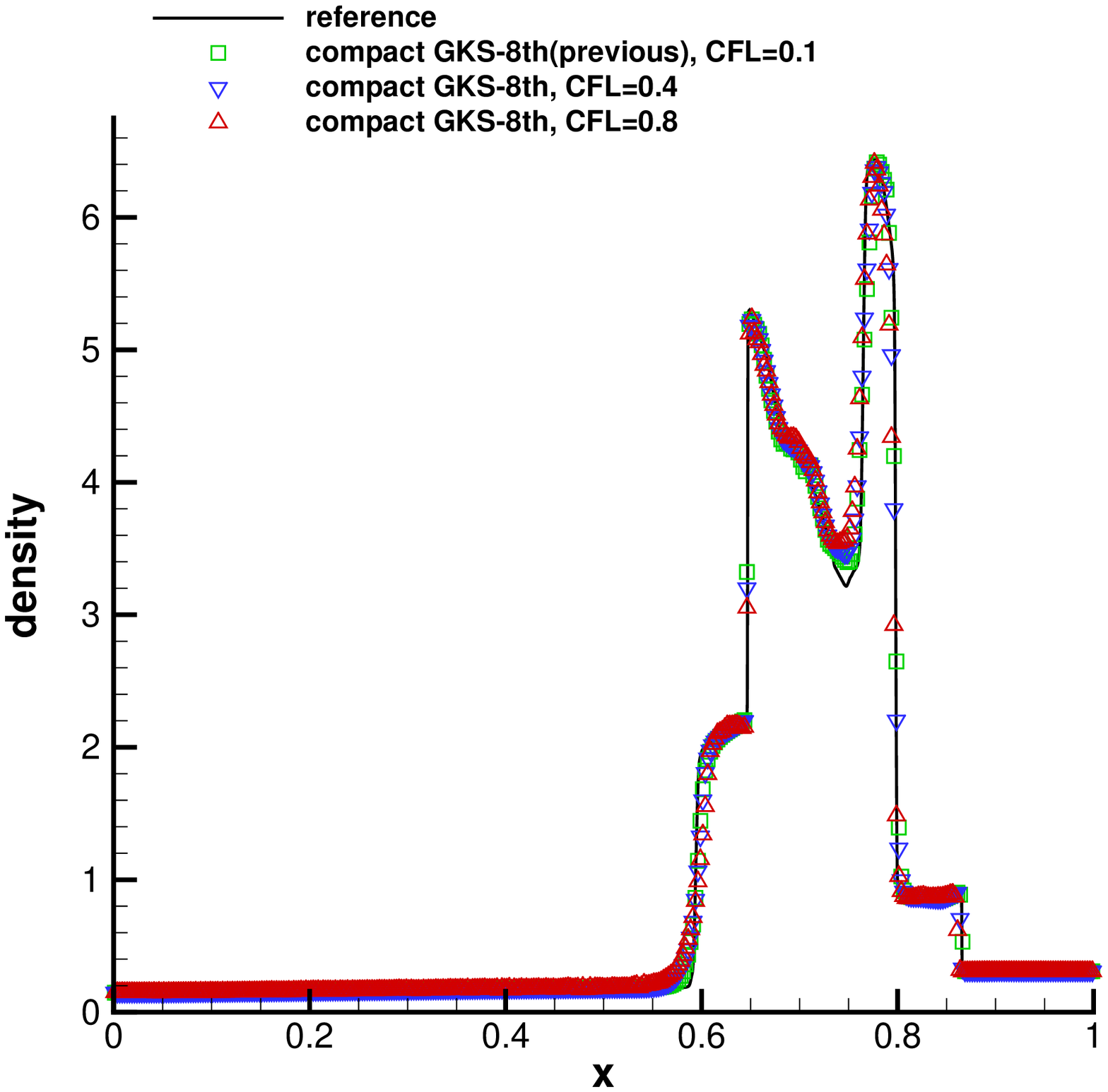}
\includegraphics[width=0.485\textwidth]{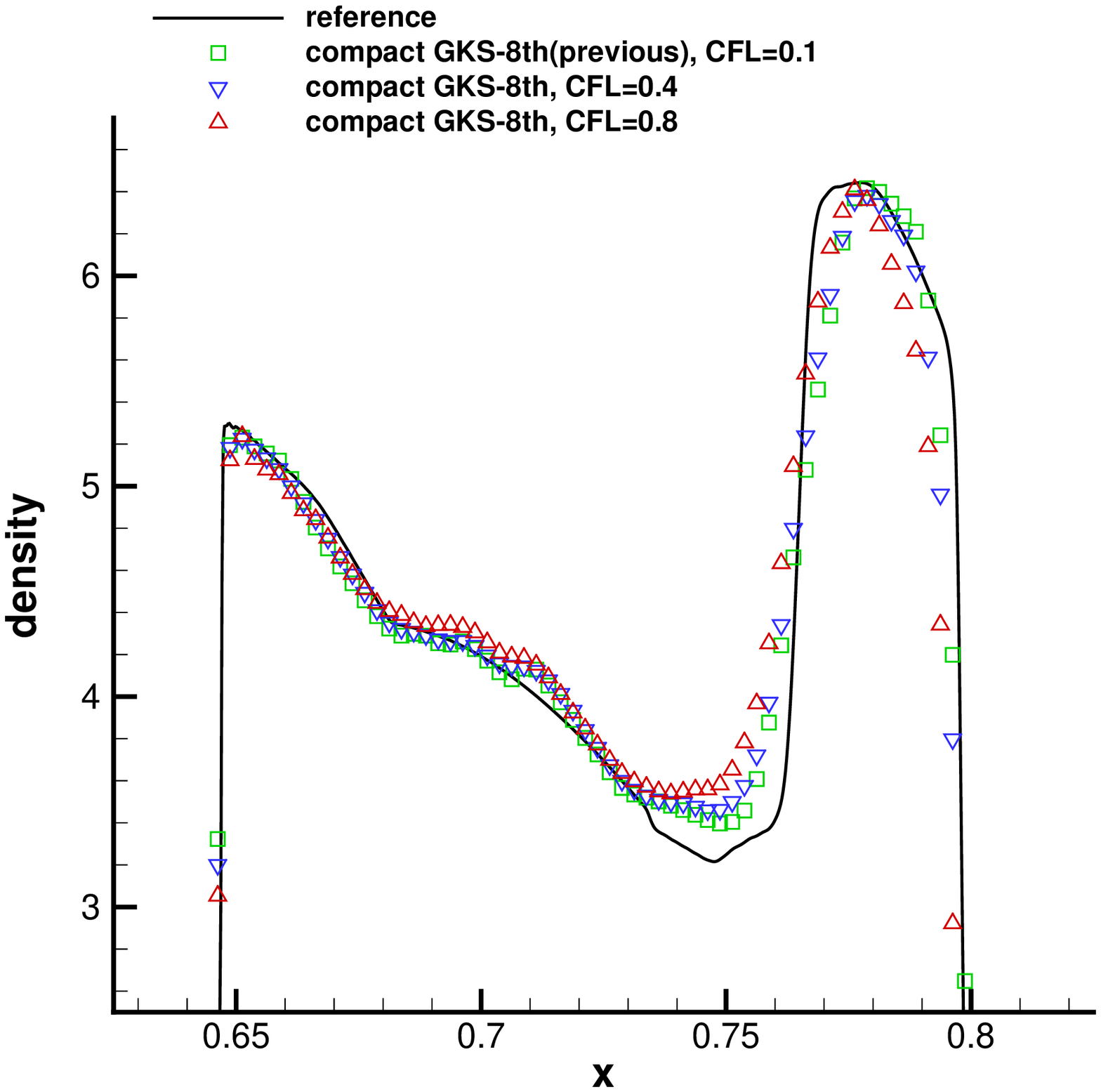}
\caption{\label{1d-Blast-wave} Blast wave problem: the density distribution at $t=0.038$  on a uniform $400$ mesh points. The CFL number takes a value $CFL=0.8$}
\end{figure}

\subsection{Blast wave problem in 1D}

To valid the scheme for the strong shock interaction, the Woodward-Colella blast wave problem  is tested \cite{Case-Woodward}.
The initial conditions is given by
\begin{equation*}
(\rho,U,p) =\left\{\begin{array}{ll}
(1, 0, 1000), \ \ \ \ & 0\leq x<0.1,\\
(1, 0, 0.01), & 0.1\leq x<0.9,\\
(1, 0, 100),  & 0.9\leq x\leq 1.
\end{array} \right.
\end{equation*}
The computational domain is $[0,1]$. The reflecting boundary conditions are imposed on both ends.
The density distributions and local enlargement from the current 8th-order compact GKS and the previous 8th-order GKS \cite{CGKSAIA} are presented in Fig. \ref{1d-Blast-wave}, where the output time is $t=0.038$ and $400$ mesh points are used.
Even based on the same reconstruction, in comparison with the previous 8th-order scheme without two flow variables update at a cell interface and limited flux function in \cite{CGKSAIA}, the current method can use a much large CFL number $0.8$ for time step, where the previous can only use
CFL number $0.2$.

\begin{figure}[!htb]
\centering
\includegraphics[width=0.485\textwidth]{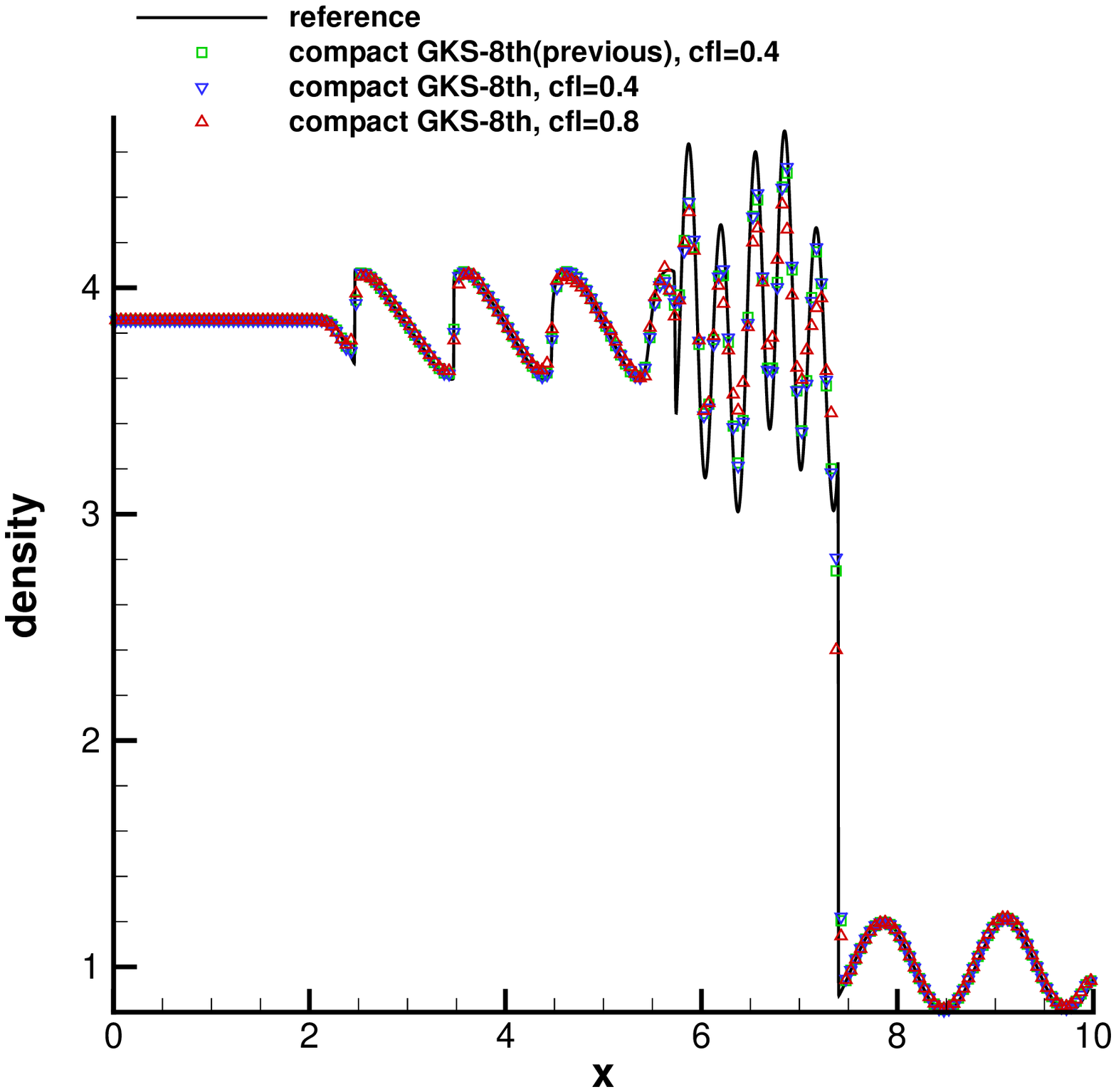}
\includegraphics[width=0.485\textwidth]{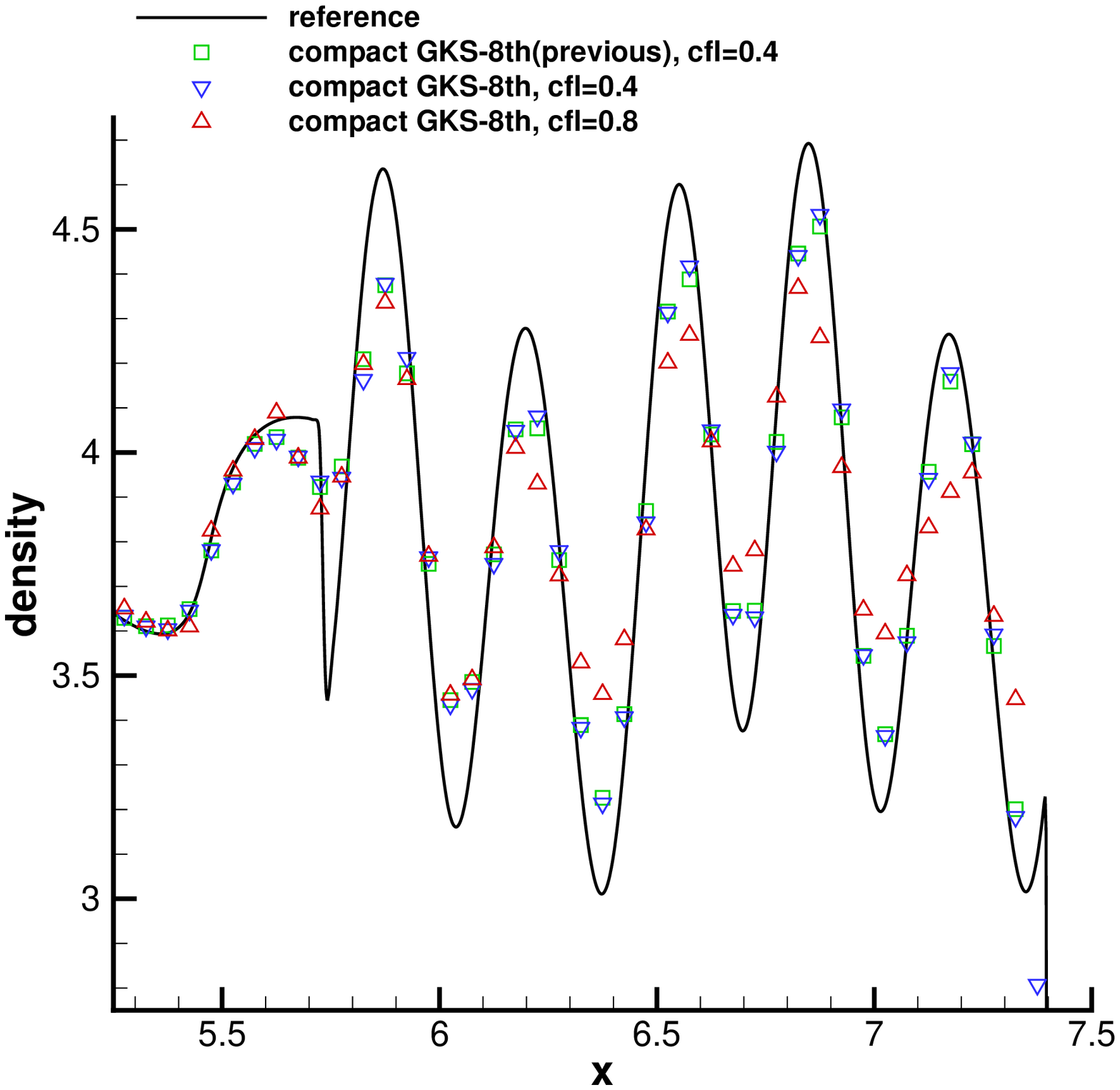}
\caption{\label{1d-Shu-osher} Shu-Osher problem: the density distributions and local enlargement at $t=1.8$ on a uniform mesh with $200$ mesh points from the current and previous 8th-order compact GKS \cite{CGKSAIA}.}
\end{figure}

\subsection{Shu-Osher problem in 1D}

The initial condition of Shu-Osher problem is \cite{Case-Shu-Osher},
\begin{equation*}
(\rho,U,p)=\left\{\begin{array}{ll}
(3.857143, 2.629369, 10.33333),  \ \ \ \ &  x \leq 1,\\
(1 + 0.2\sin (5x), 0, 1),  &  1 <x \leq 10.
\end{array} \right.
\end{equation*}
The computational domain is $[0, 10]$ and $200$ uniform mesh points are used.
The non-reflecting boundary condition is used at both ends.
The density profiles and local enlargement at $t=1.8$ from the current and previous 8th-order GKS are shown in Fig. \ref{1d-Shu-osher}.
A large CFL number can be used in the current calculation.

\begin{figure}[!htb]
	\centering
	\includegraphics[width=0.495\textwidth]{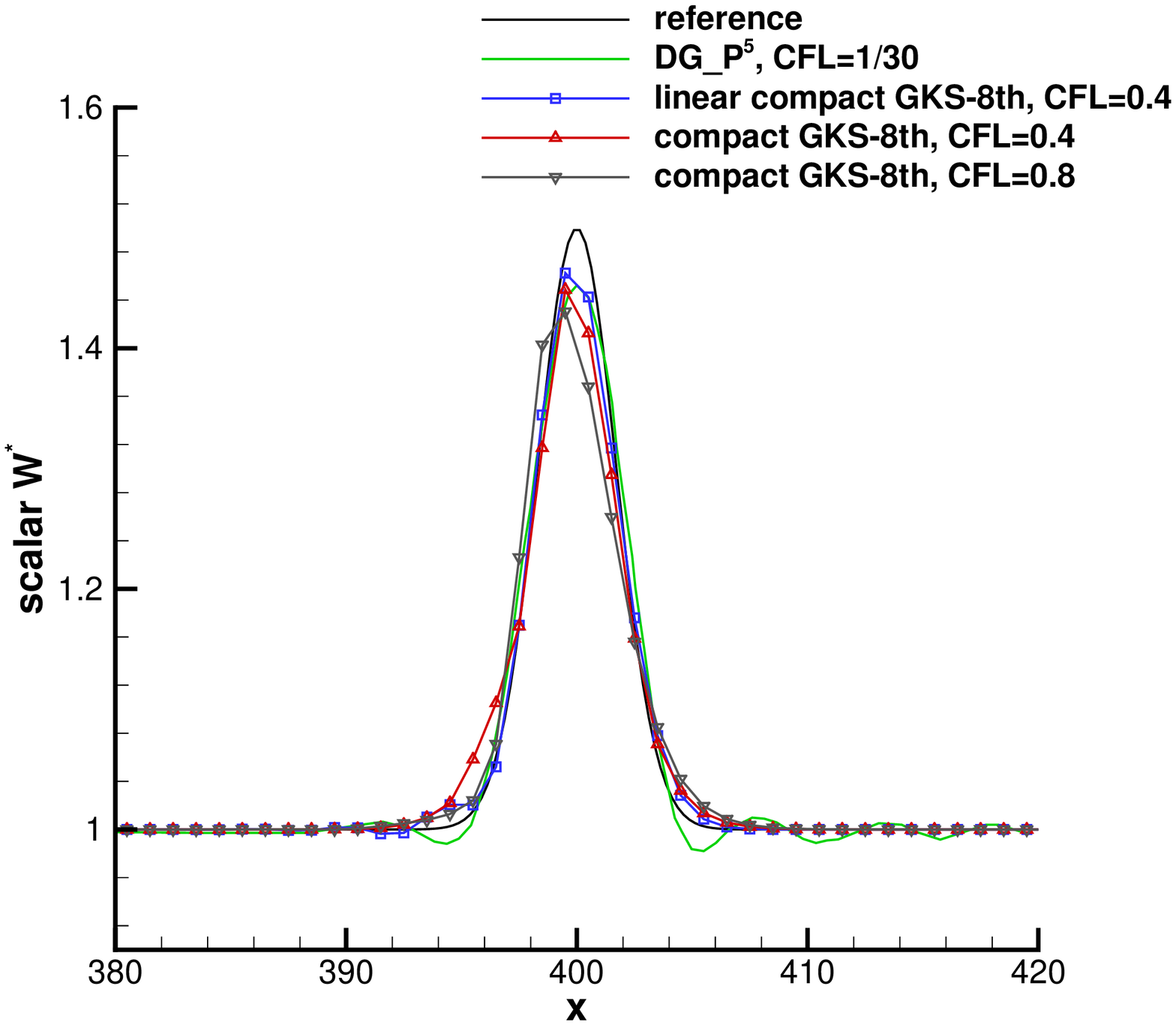}
	\includegraphics[width=0.495\textwidth]{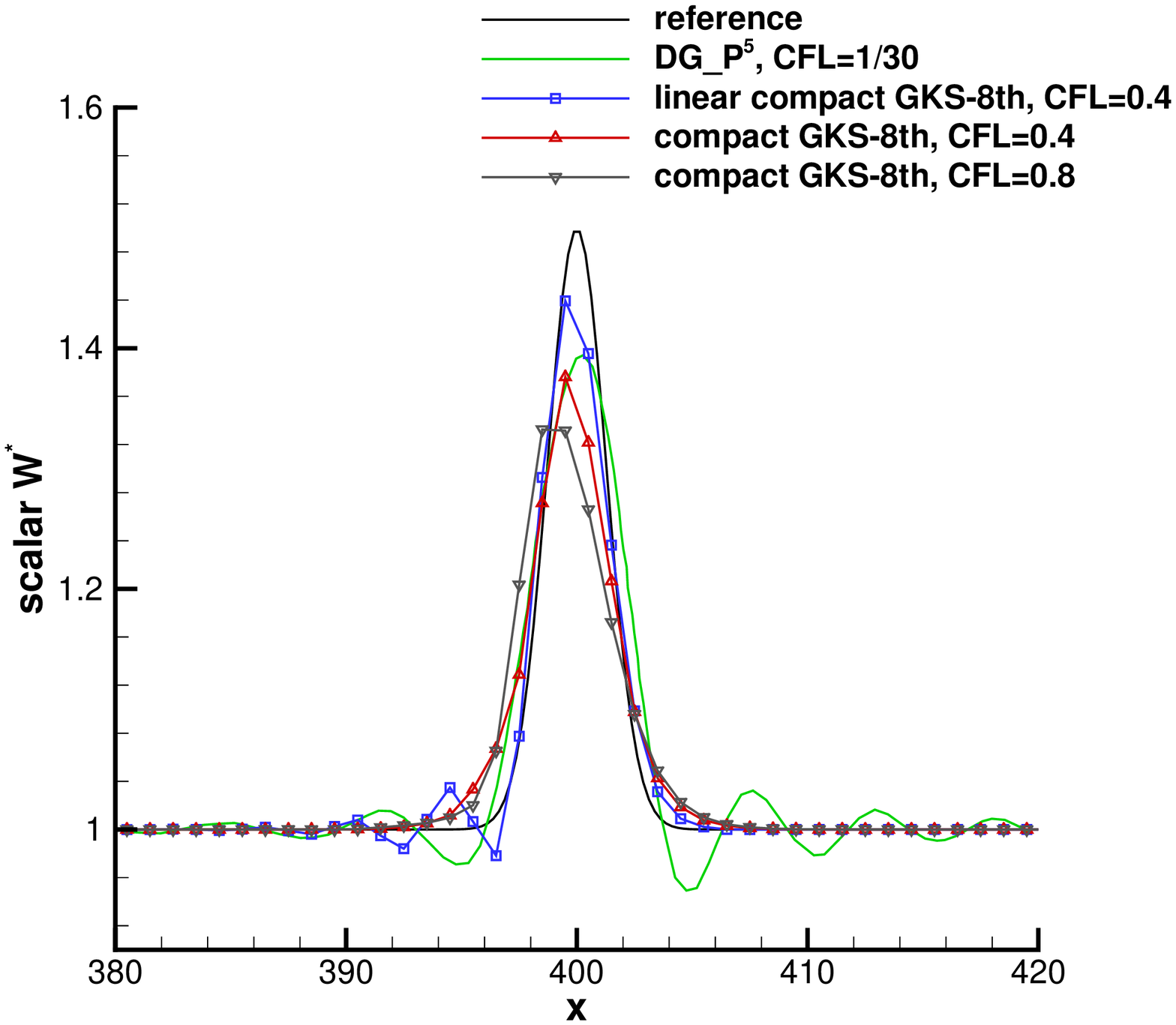}
	\caption{\label{1d-advection} Linear advection problem: the results of 8th-order compact GKS at $t=400$ with $b=2.0$ and $b=1.5$. The mesh size is $1$.  }
\end{figure}

\subsection{Advection of entropy wave in 1D}
The test is a problem of advection of 1-D entropy wave in a stationary mean flow. The initial condition is defined as
\begin{align*}
&\rho= 1 + 0.5 e^{-\ln 2 x^2/b^2},\\
& p=1, \\
& U=1,
\end{align*}
where different $b=2.0$ and $b=1.5$ are tested respectively. The computational domain is $[-800,1000]$. Free flow boundary condition is adopted.
The similar test has been presented in \cite{DGacoustic2019}, where the linear advection equation is solved directly and a very small CFL number $(CFL<0.05)$ is adopted for their high-order finite difference and DG schemes.
The initial value problem solved in \cite{DGacoustic2019} has no a background uniform flow distribution, but the solution is consistent
with the test in this paper.

Fig. \ref{1d-advection} shows the results of compact GKS with different CFL numbers and reconstructions. The cell size of the uniform mesh is $h=1$. The only difference between the linear compact GKS-8th and the compact GKS-8th is that the linear reconstruction is used in the linear case,
which is achieved simply by replacing the nonlinear weights with the linear ones.
At the same time, one of the best results listed in \cite{DGacoustic2019}, which comes from the sixth-order DG-$P^5$ scheme,
is used for comparison. For the DG-$P^5$ scheme, there are six independently evolution equations in each mesh cell, the cell size used in the solution of the DG-$P^5$  is $h=6$, and the CFL number is $1/30$.
In the compact GKS, there is only one evolution equation for the gas distribution function at a cell interface for the updates of cell-averaged values and cell-averaged gradients. The CFL numbers in the GKS are CFL$=0.4$ and $0.8$.
The output time step is dt$=0.2$.
In comparison with the results of DG-$P^5$ scheme, the current scheme works very well even for the linear wave propagation.

\subsection{Blast wave problem in 2D on triangular mesh}

\begin{figure}[!htb]
	\centering
	\includegraphics[width=0.475\textwidth]{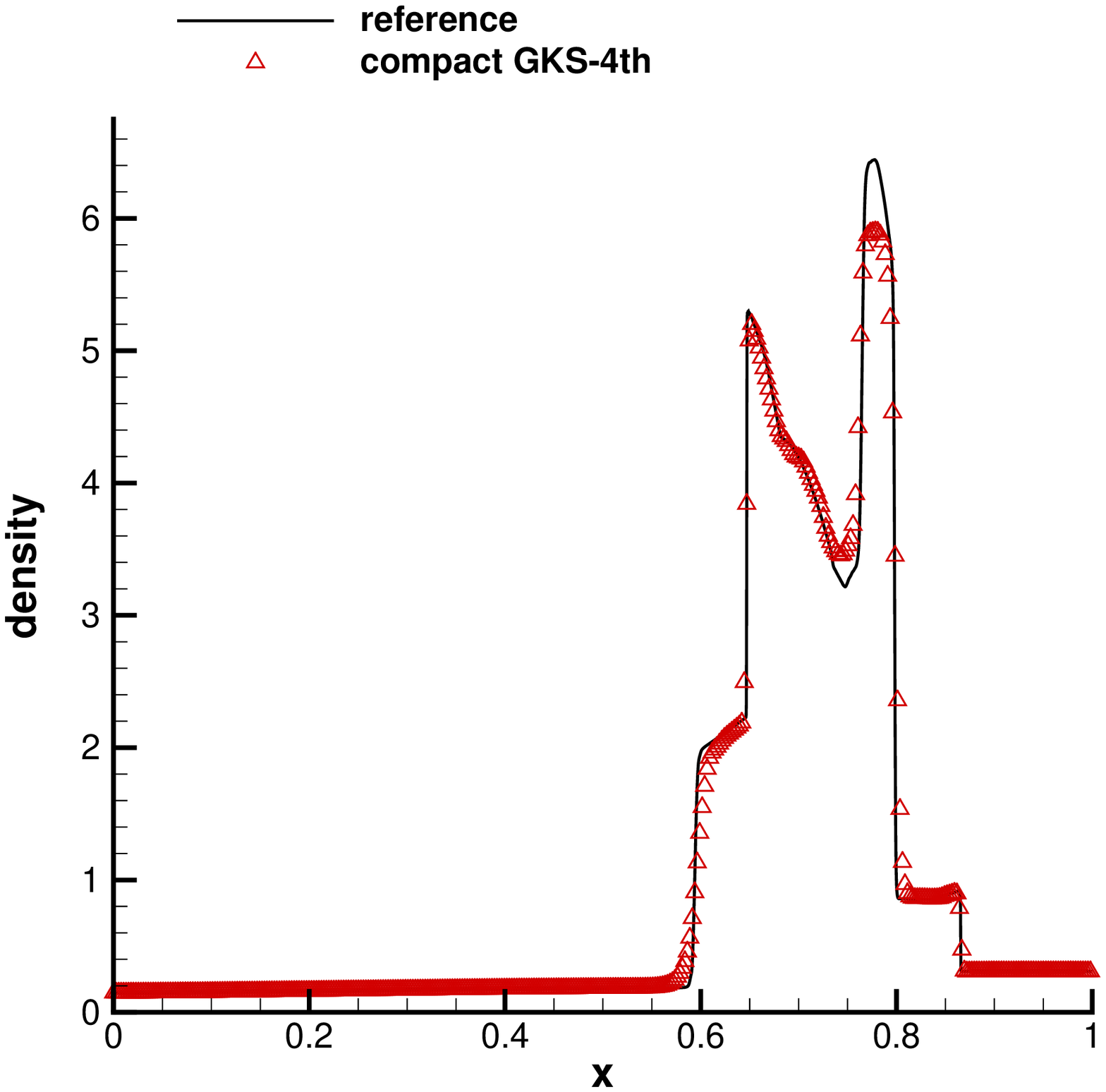}
    \includegraphics[width=0.475\textwidth]{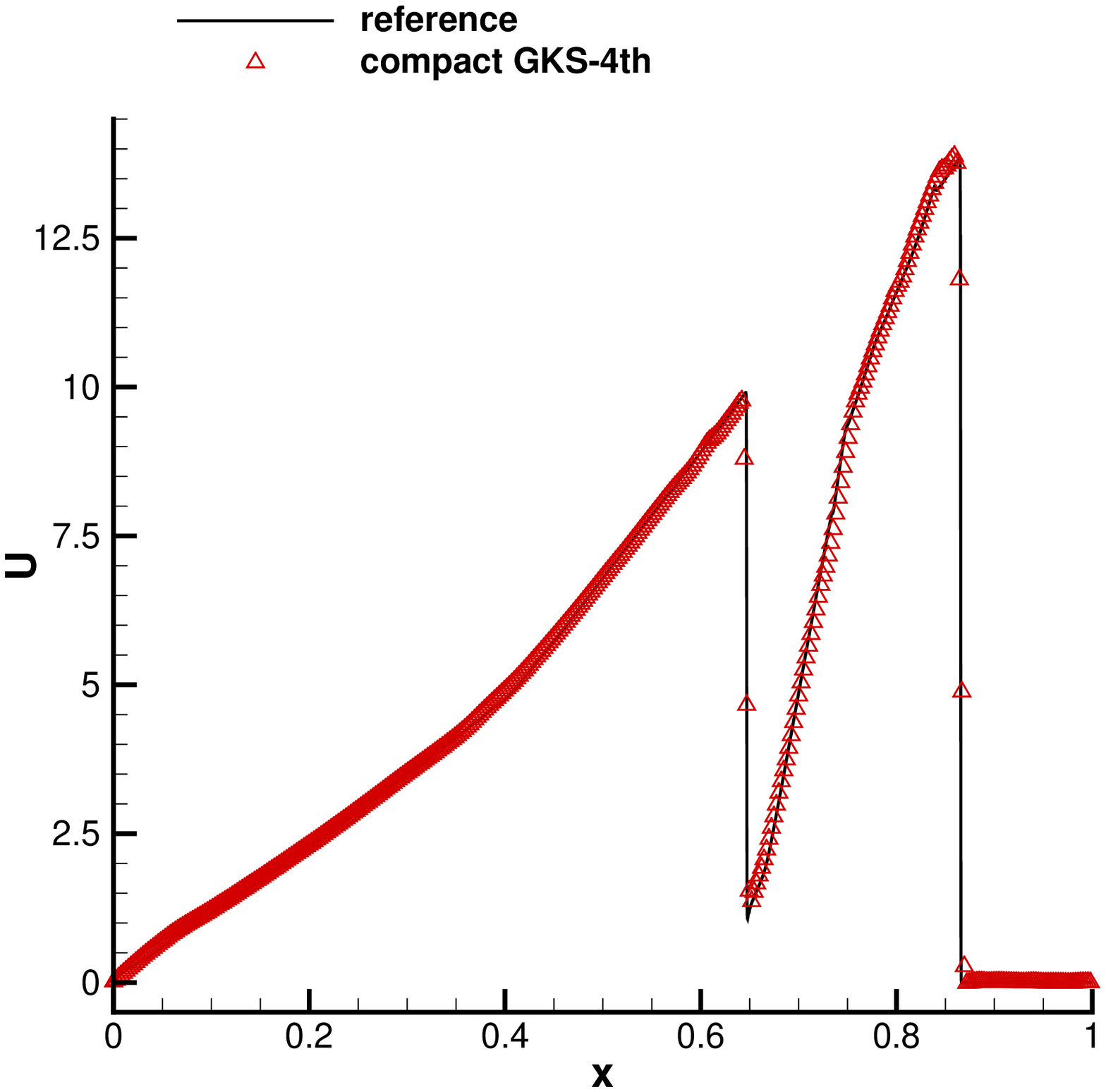}
	\caption{\label{2-blastwave} Blast wave problem: the density and velocity distributions along the horizontal centerline by the 4th-order compact GKS on triangular mesh at $t=0.038$ with cell size $h=1/400$. The CFL number takes a value $CFL=0.8$. }
\end{figure}

To test the compact scheme  on unstructured mesh, the Woodward-Colella blast wave problem is tested again \cite{Case-Woodward}.
The computational domain is $[0,1]\times[0,0.25]$, and the reflecting boundary conditions are imposed on all boundaries. The triangular mesh with $h=1/400$ is used. In this test case, the $CFL$ number also takes $0.8$. The computed density and velocity profiles at $t=0.038$ along the central horizontal line of the computational domain are shown in Fig. \ref{2-blastwave}.

\subsection{Mach 3 step problem on triangular mesh}

\begin{figure}[!htb]
	\centering
	\includegraphics[width=1.0\textwidth]{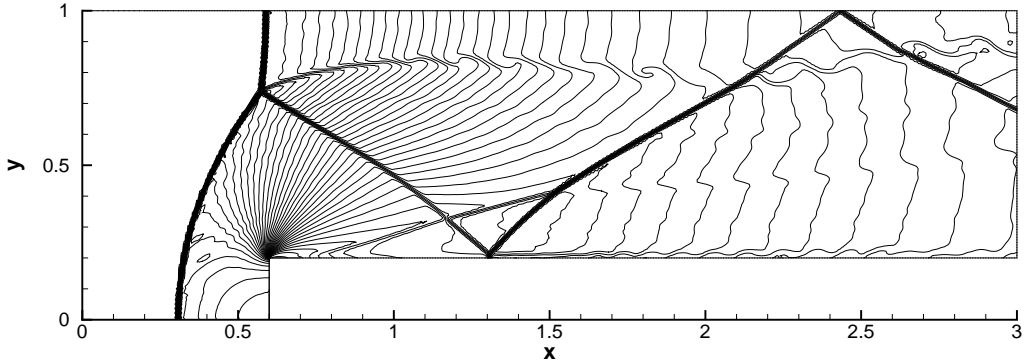}
	\caption{\label{2-forward-step} Forward step problem: the density by 4th-order compact GKS on triangular mesh at $t=4.0$ with cell size $h=1/120$. }
\end{figure}

The step problem was extensively studied in \cite{Case-Woodward} for inviscid flow. The computational domain is $[0,3]\times[0,1] \setminus [0.6,3]\times[0,0.2]$. The height of the wind tunnel is $1$, and the length is $3$. The step is located at $x=0.6$ with height $0.2$ in the tunnel. The gas in the tunnel has an initial condition $\rho=1, U=3, V=0, p=1/1.4$.
The same state is used as the left boundary condition.
The upper and lower boundaries are imposed with slip Euler boundary condition.
The corner of the step is the center of a rarefaction fan. The method of modifying the density and velocity magnitude on the several cells around the corner has not been used in the current computation \cite{Case-Woodward}. The local solution at the corner is properly resolved by the compact GKS on the regular mesh with mesh size $1/120$.
The density distribution at time $t=4.0$ is plotted in Fig. \ref{2-forward-step}. The current scheme provides a high resolution solution, such as the
capturing of the physical instability around the slip line. At the same time,
excellent shock capturing capability of the compact scheme has been validated.

\subsection{Hypersonic flow around a cylinder on unstructured mesh}

\begin{figure}[!htb]
	\centering
    \includegraphics[width=0.30\textwidth]{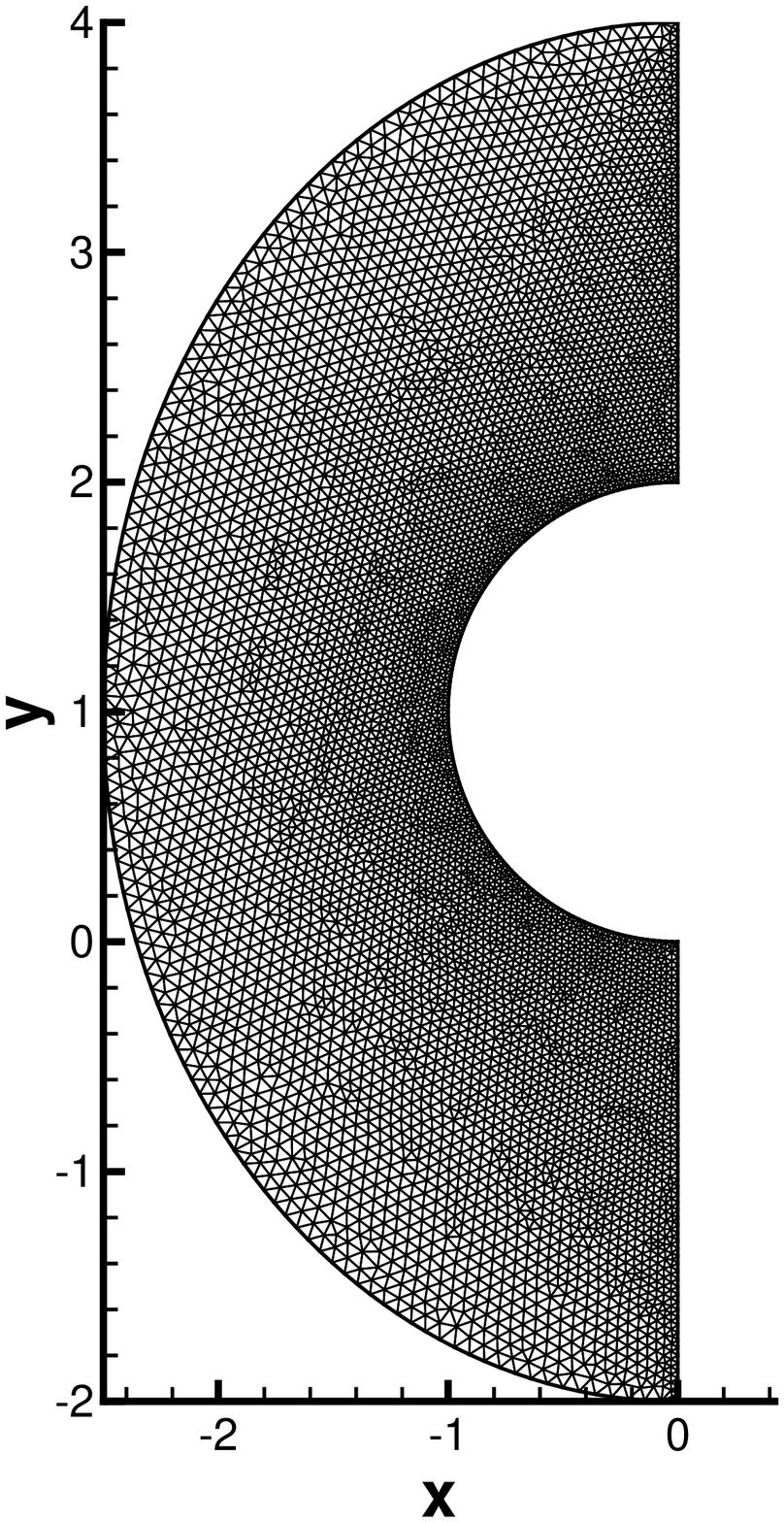}
	\includegraphics[width=0.30\textwidth]{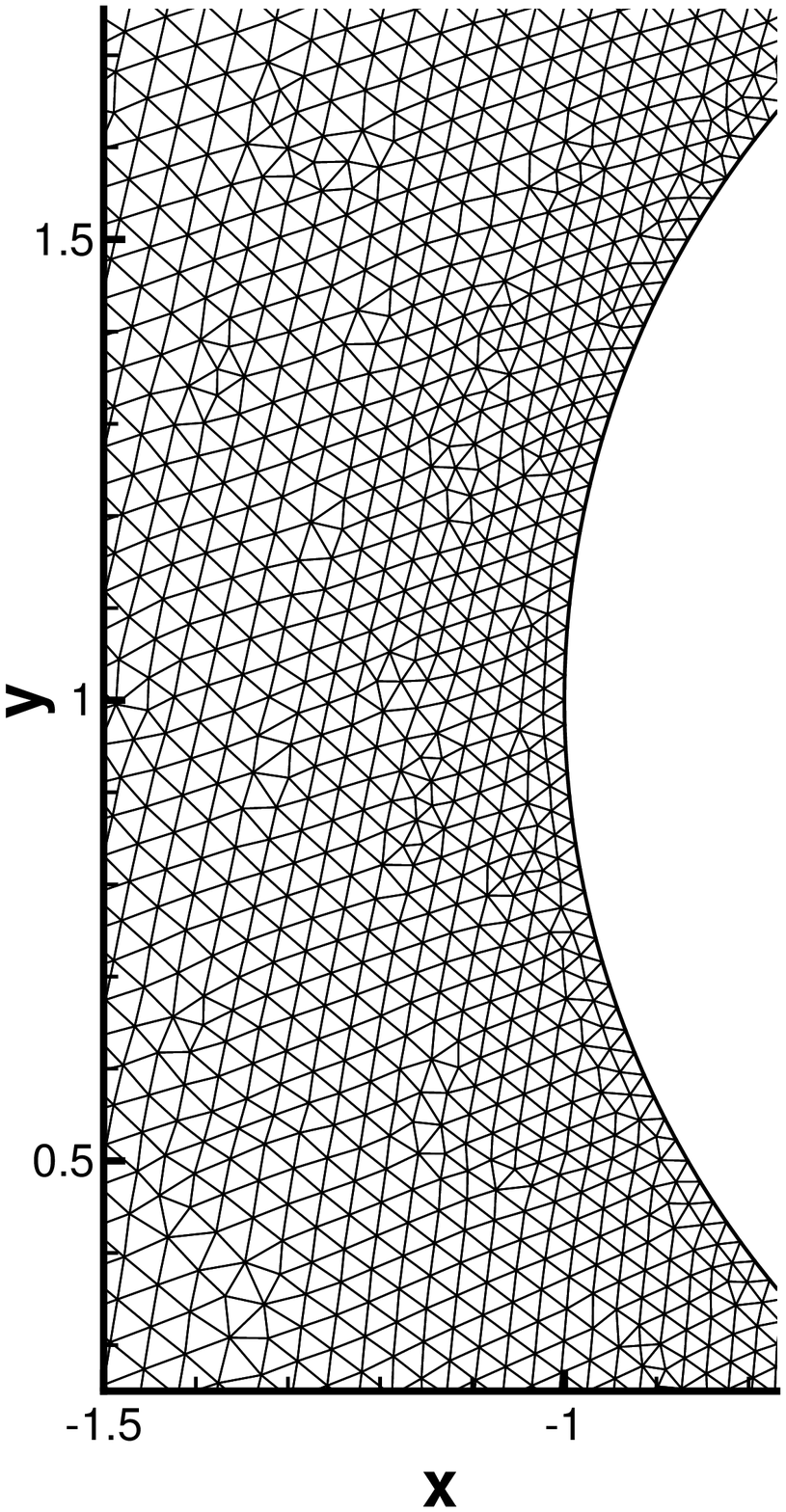}
	\caption{\label{2-inviscid-cylinder-1} Inviscid cylinder flow at mach 20: the computational mesh and the local enlargement.}
\end{figure}

The incoming inviscid flow has a Mach number $8$ and $20$ separately. The adiabatic reflective boundary condition is imposed on the wall of the cylinder.
The mesh and pressure distributions are presented in Fig. \ref{2-inviscid-cylinder-1}. The regular triangular mesh is used, and the mesh is refined near the cylinder. The results agree well with those calculated on structured mesh by the non-compact high-order GKS.
The CFL number used in the calculation is CFL$=0.8$ and the 4th-order compact GKS can work well for Mach $20$ flow starting at $t=0$ without additional treatment in the initial phase.

\begin{figure}[!htb]
	\centering
    \includegraphics[width=0.35\textwidth]{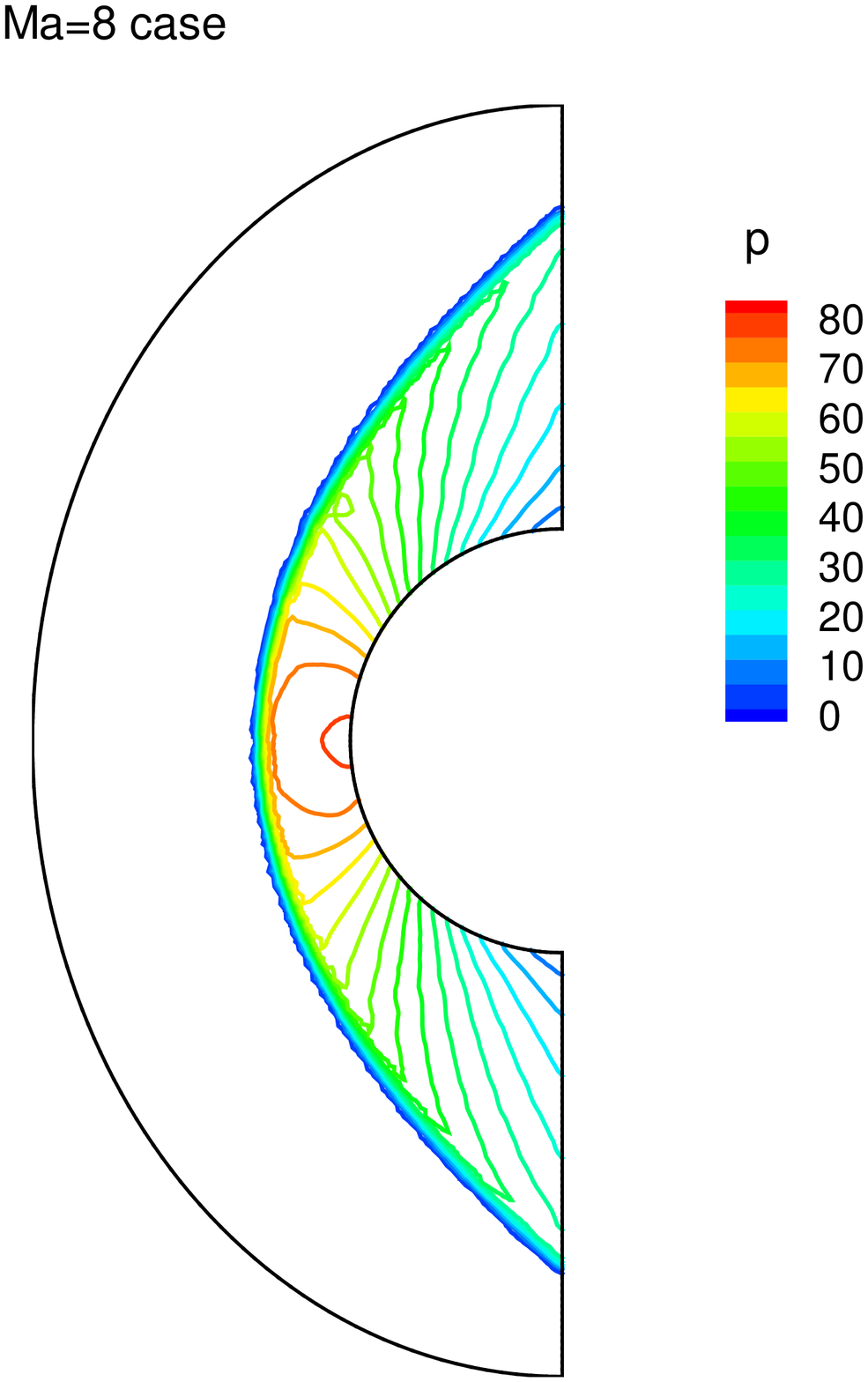}
    \includegraphics[width=0.35\textwidth]{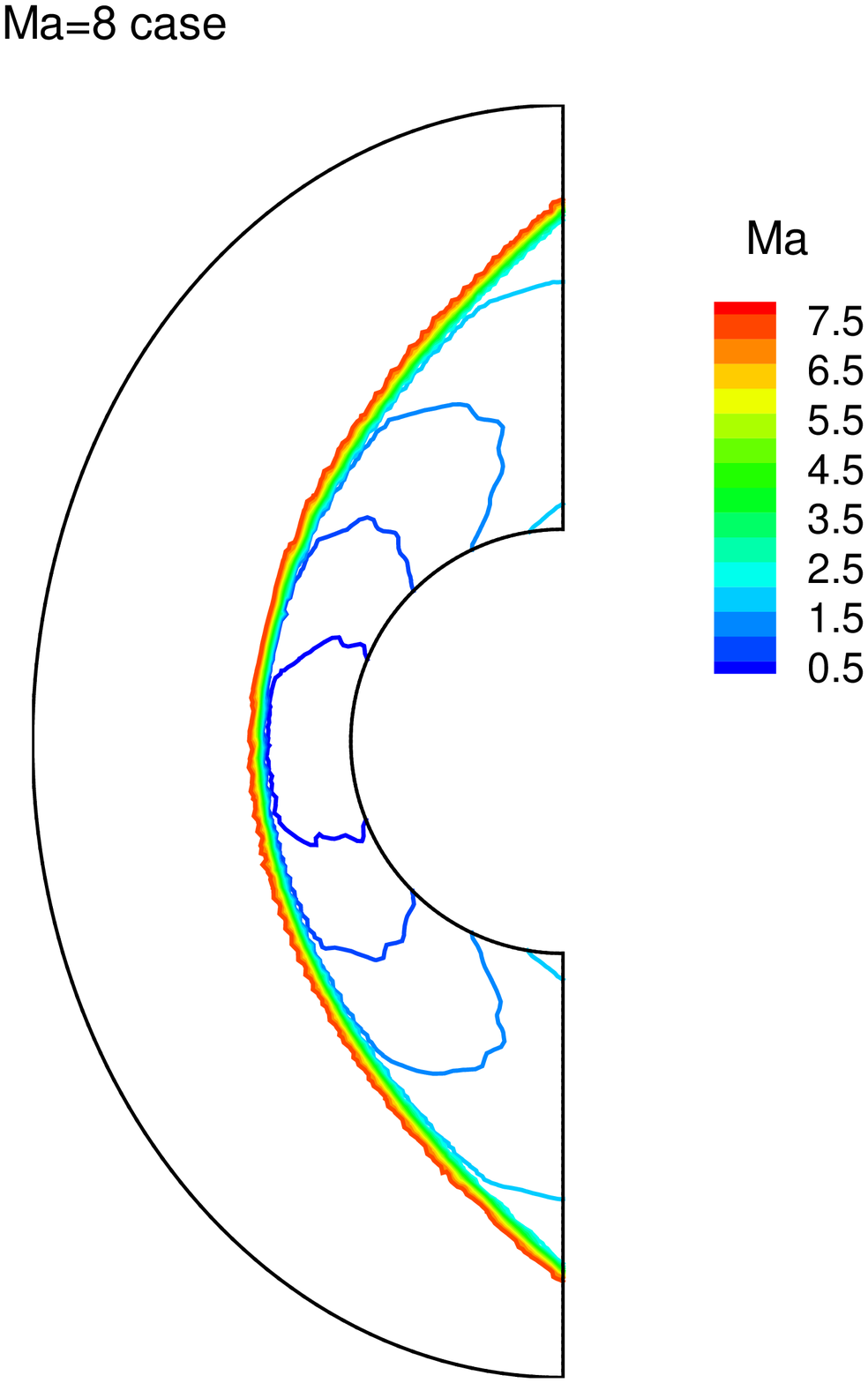}\\
    \includegraphics[width=0.35\textwidth]{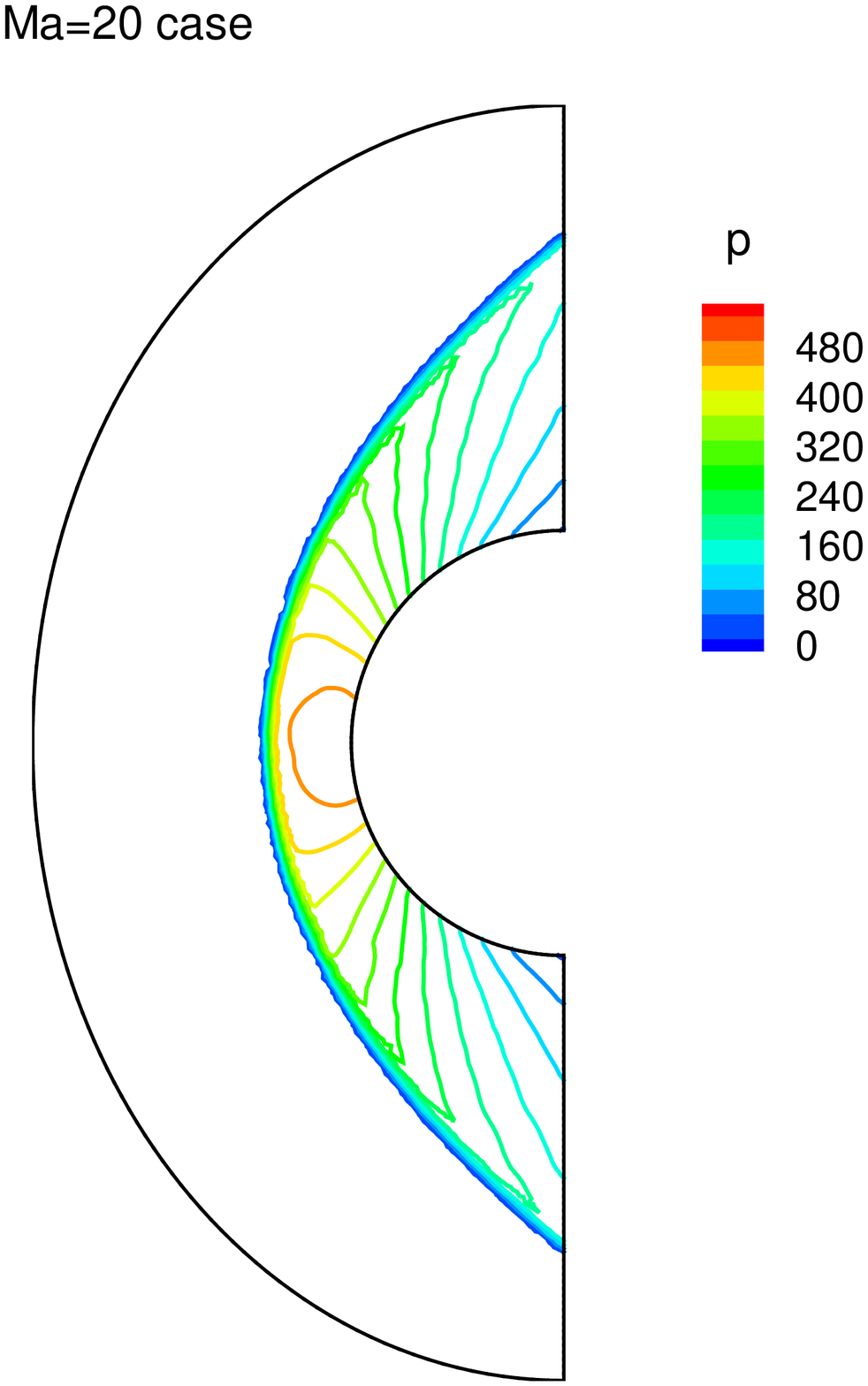}
    \includegraphics[width=0.35\textwidth]{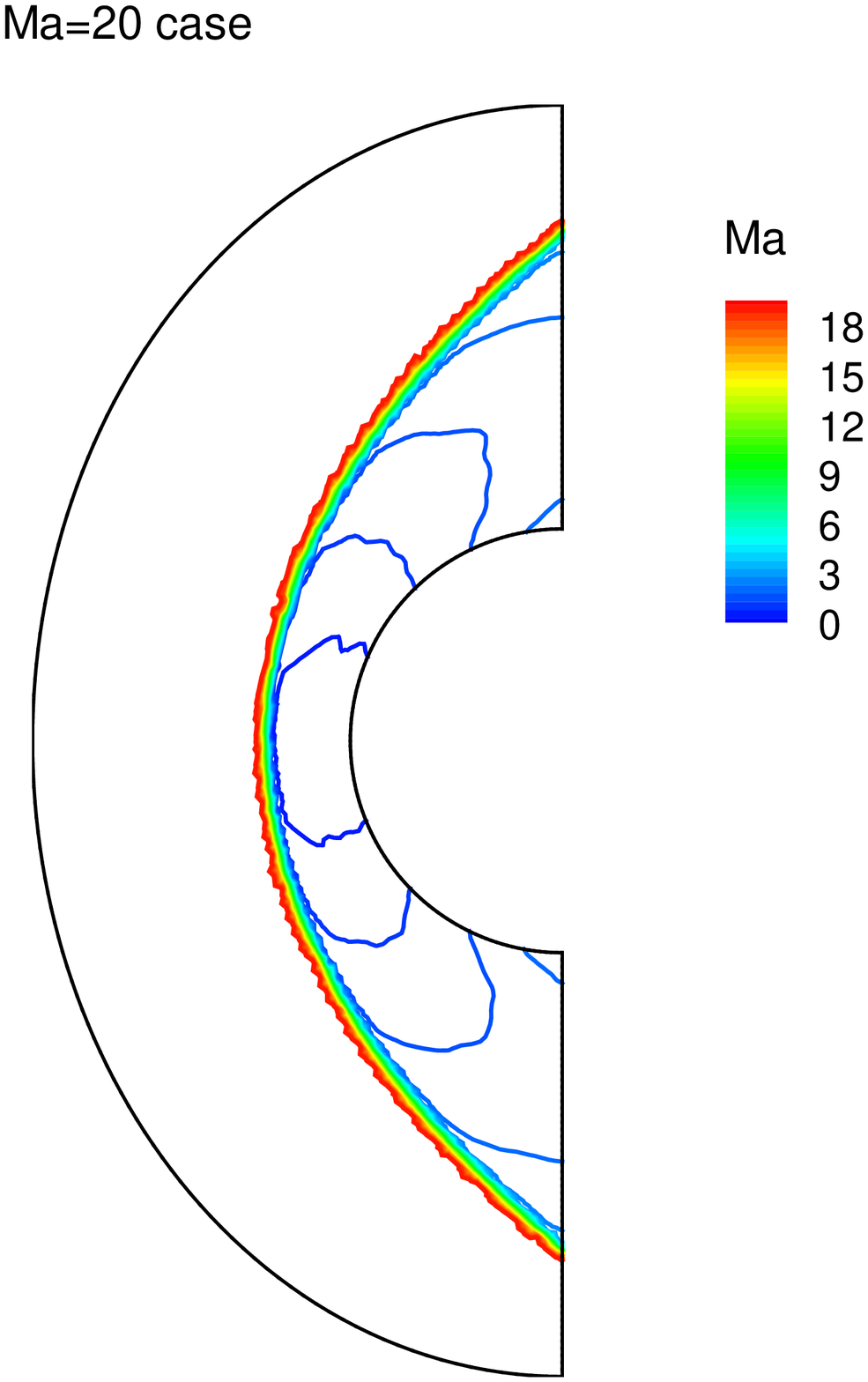}
	\caption{\label{2-inviscid-cylinder-2} Inviscid cylinder flow at mach 20: pressure and Mach number distributions of $Ma=8$ and $20$ cases by the 4th-order compact GKS on triangular mesh. CFL number takes a value CFL$=0.8$}
\end{figure}

\begin{figure}[!htb]
	\centering
    \includegraphics[width=0.325\textwidth]{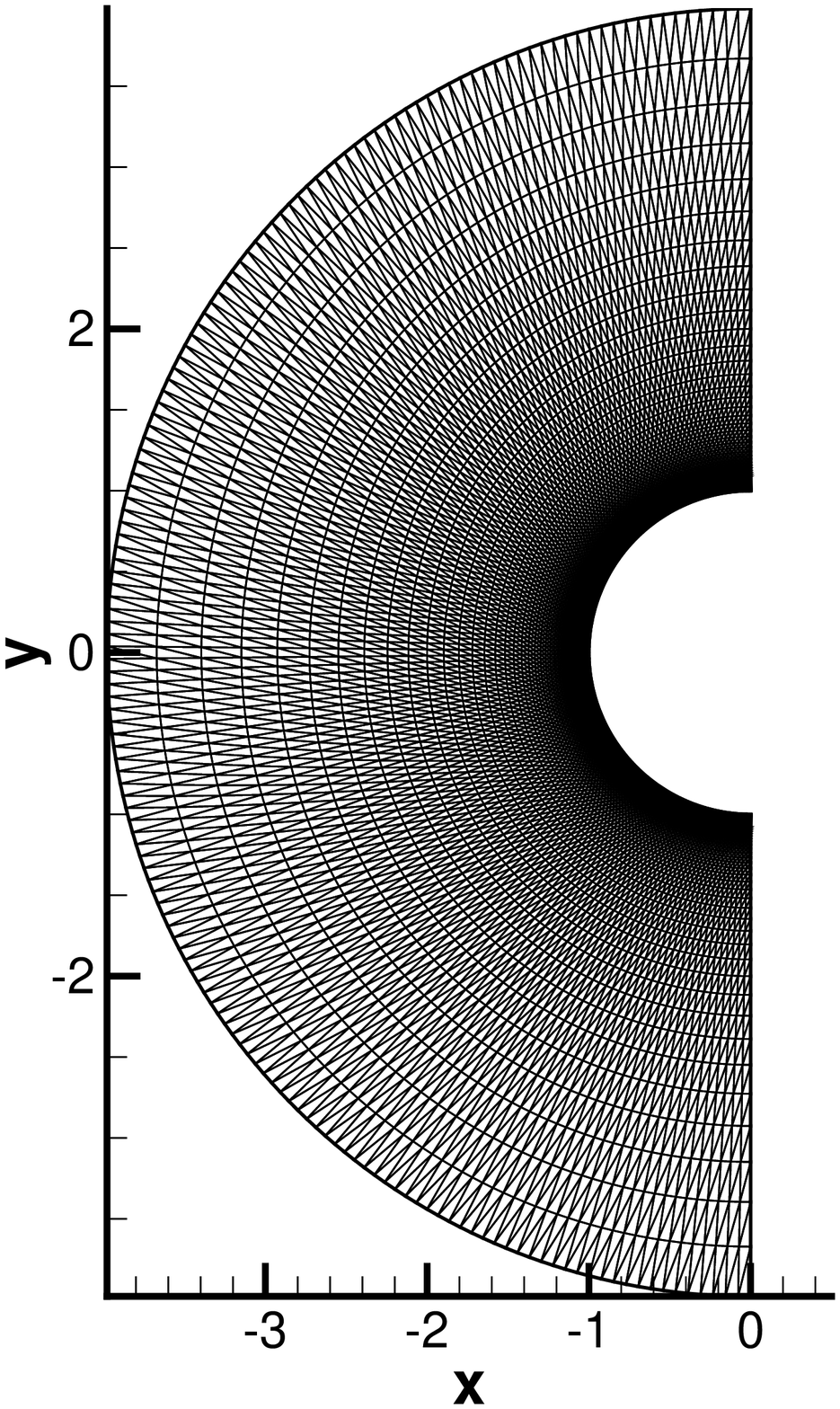}
    \includegraphics[width=0.325\textwidth]{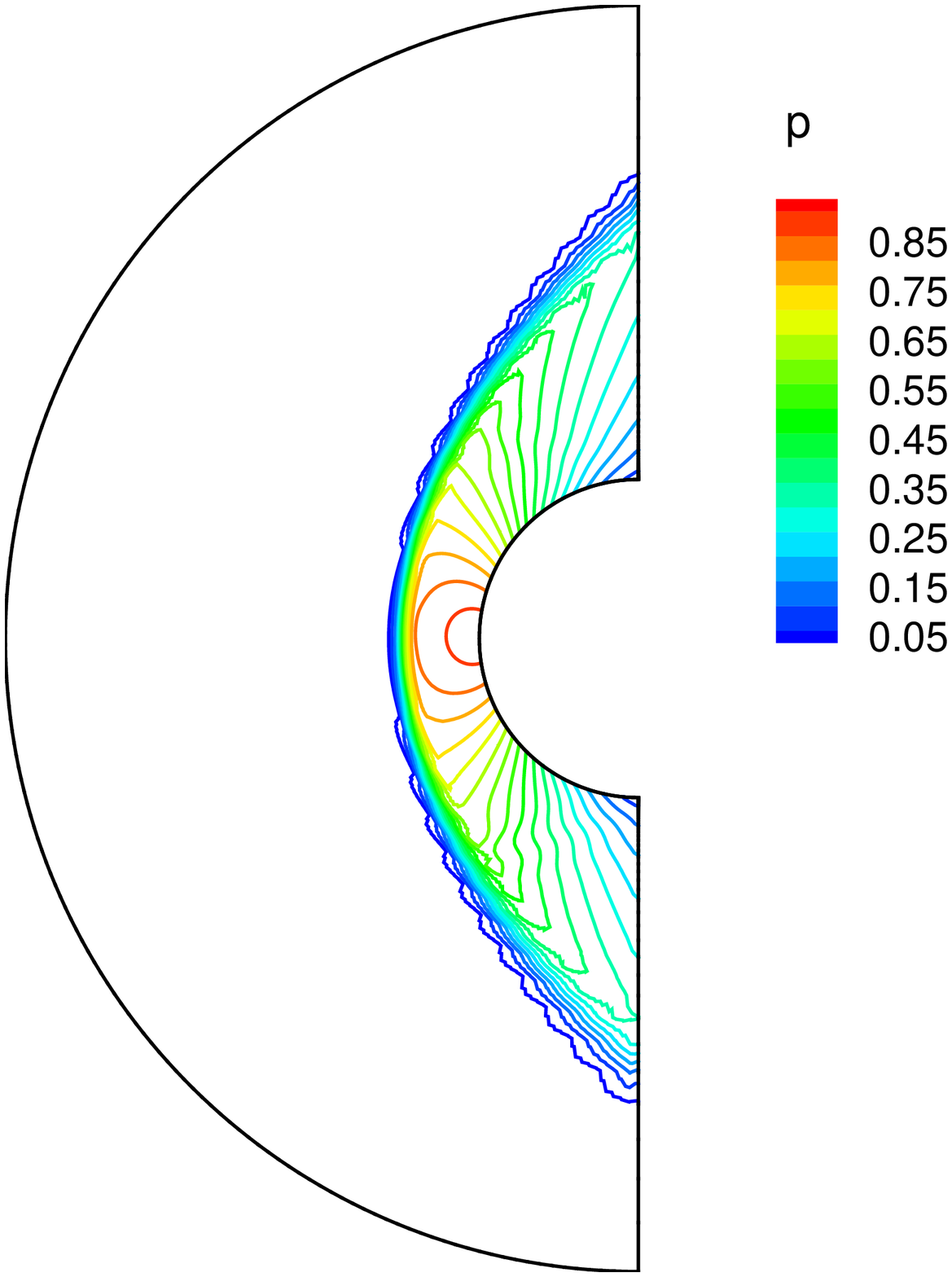}
    \includegraphics[width=0.325\textwidth]{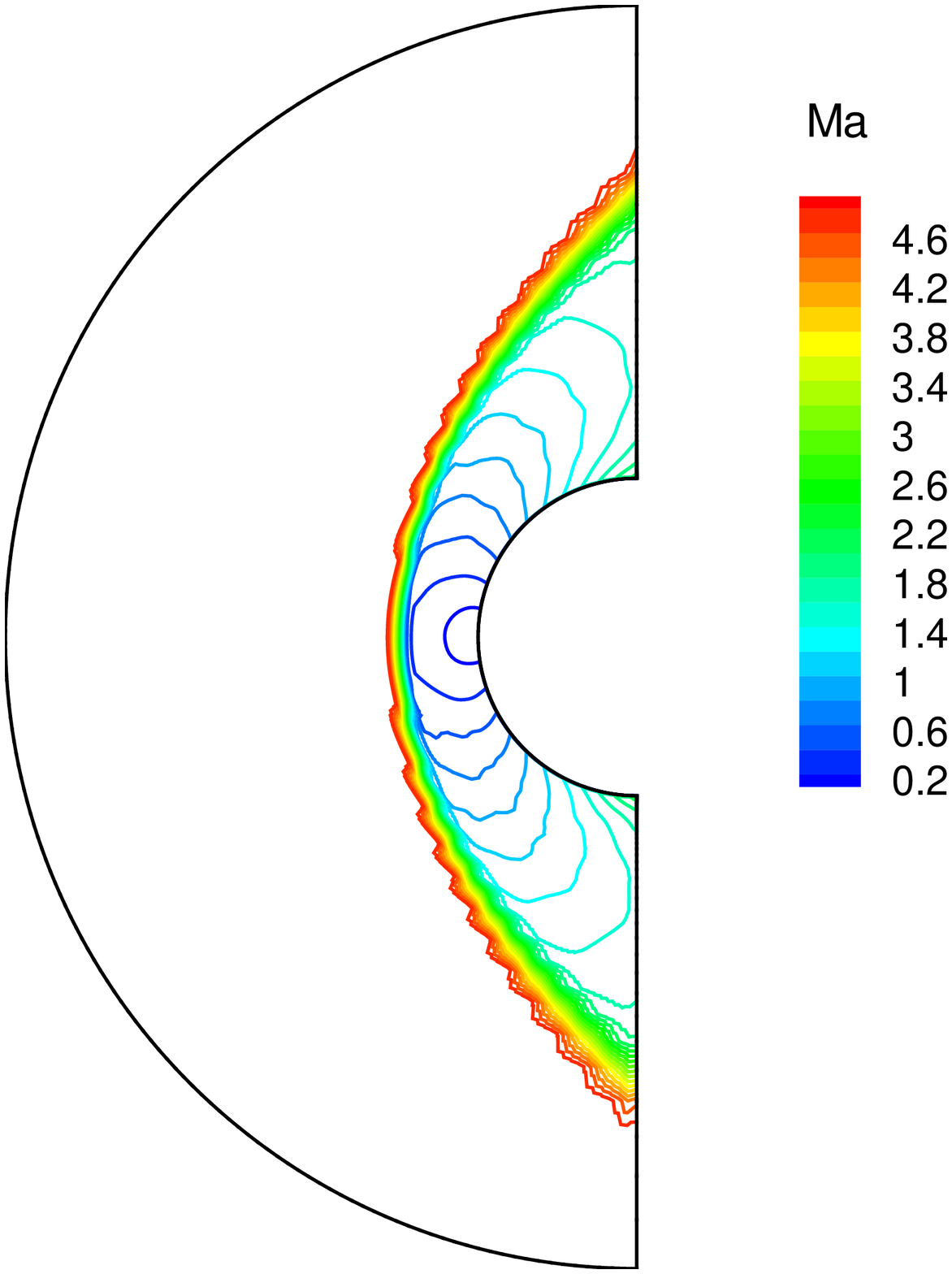}
	\caption{\label{2-viscous-cylinder-1} Viscous flow around a cylinder. The pressure and Mach number distributions at $Ma=5$ by 4th-order compact GKS on triangular mesh.}
\end{figure}

The scheme is further tested for viscous flow computation at a Mach number $5$ on triangular mesh with a large aspect ratio.
The Mach $Ma=5$ incoming flow has a temperature $T_{\infty}=124.94K$. The Reynolds number is $Re=1.835\times 10^5$.
The isothermal non-slip wall boundary condition is imposed on the surface of the cylinder with a wall temperature $T_w=294.44K$.
The right boundary is set as outflow boundary condition.
Because of the viscous flow, a CFL number CFL=$0.35$ is used in the calculation. The mesh and flow distributions are shown
in Fig.\ref{2-viscous-cylinder-1}.

\subsection{Viscous shock tube on triangular mesh}

\begin{figure}[!htb]
	\centering
    \includegraphics[width=0.495\textwidth]{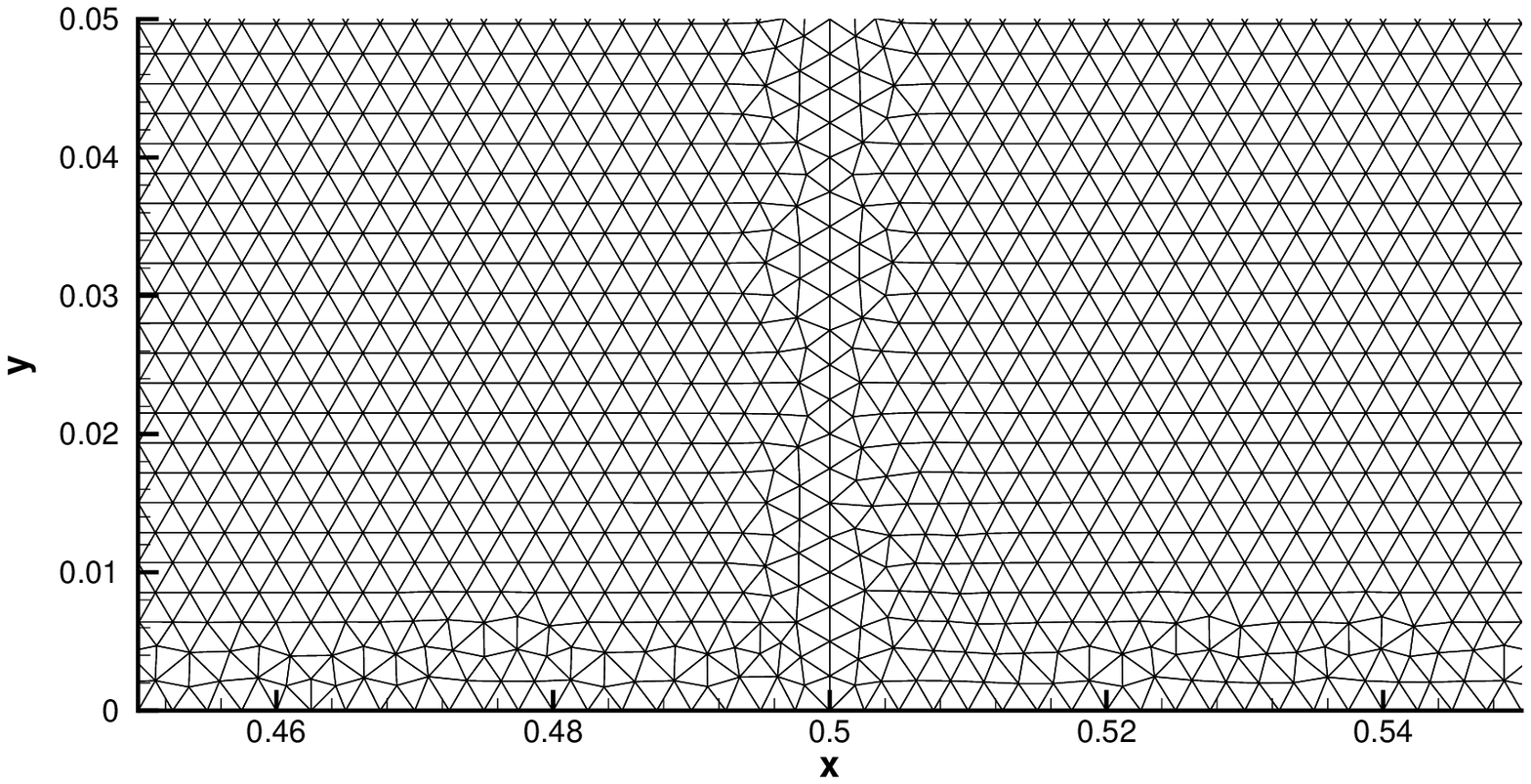}
	\includegraphics[width=0.495\textwidth]{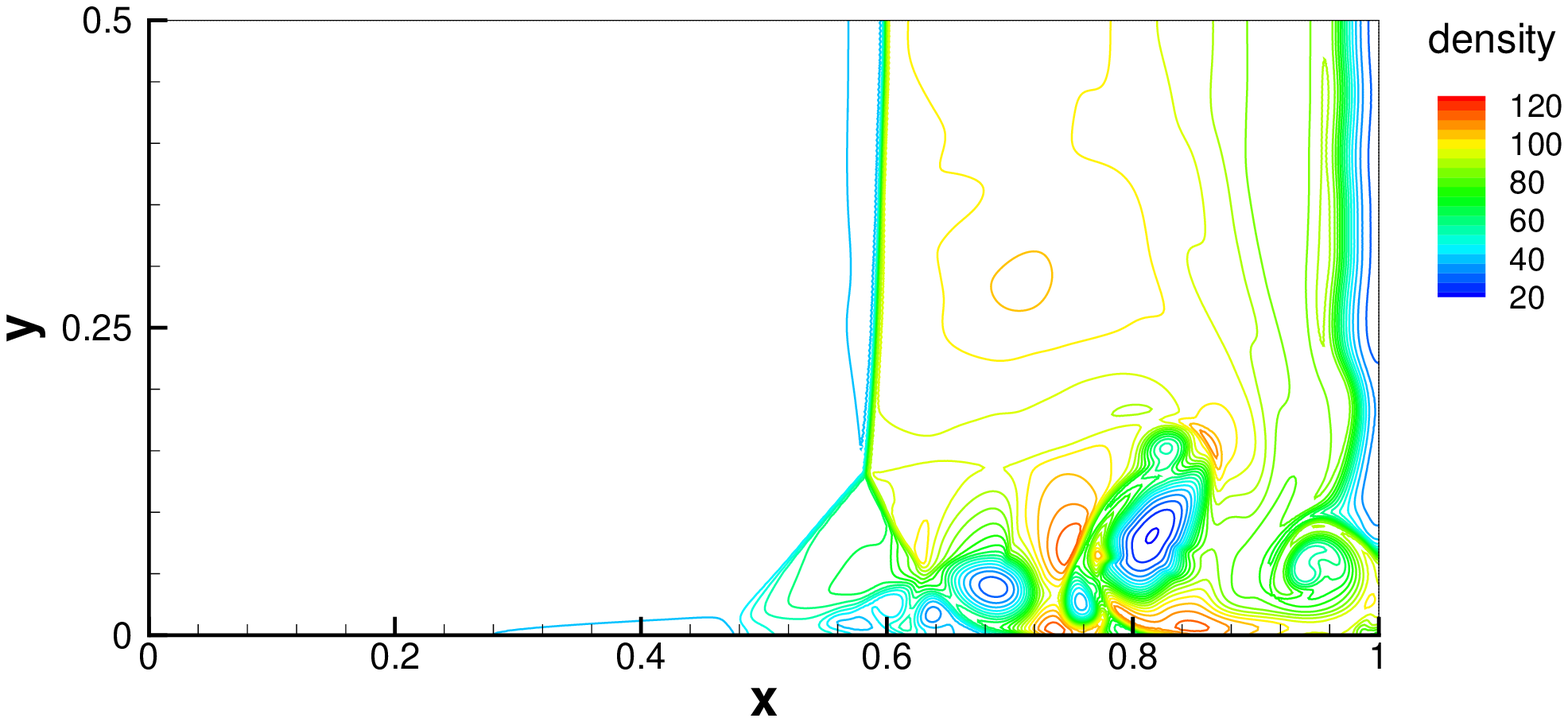}
	\caption{\label{2-viscous-tube-1} Viscous shock tube flow. Local computational mesh with cell size $h=1/400$ and the density contours.}
\end{figure}

\begin{figure}[!htb]
	\centering
    \includegraphics[width=0.625\textwidth]{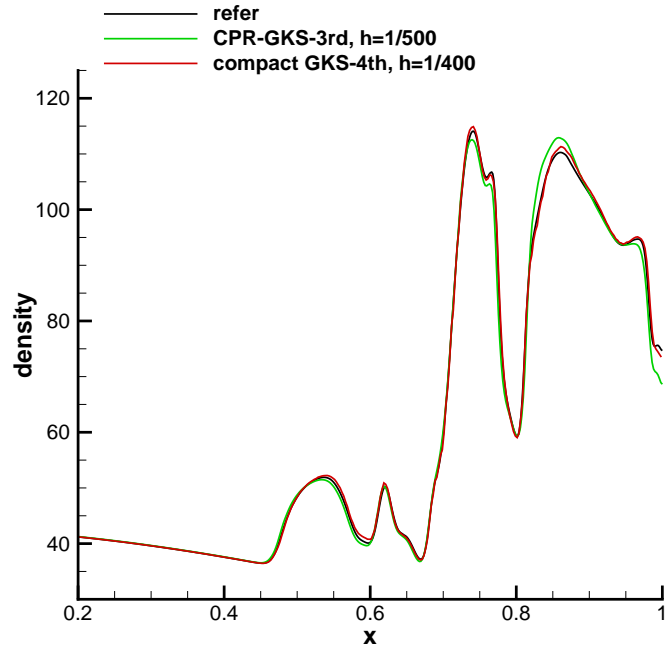}
	\caption{\label{2-viscous-tube-2} Viscous shock tube flow. The density distribution along the lower wall.}
\end{figure}

Viscous shock tube problem is the computation of complicated flow phenomena associated with shock wave and boundary layer interaction.
This problem requires not only the robustness of the scheme, but also the accuracy of the numerical method. The flow is bounded in a unit square cavity. The computational domain is set as $[0,1] \times [0,0.5]$ and a symmetrical boundary condition is used on the top boundary.
The non-slip and adiabatic wall conditions are imposed on other boundaries. The initial condition is
\begin{equation*}
(\rho,U,V,p) = \begin{cases}
(120,0,0,120/\gamma),  0\leq x<0.5,\\
(1.2,0,0,1.2/\gamma),  0.5\leq x\leq1.
\end{cases}
\end{equation*}
The viscosity coefficient is $\mu=0.005$ with a corresponding Reynolds number $Re=200$.
The Prandtl number in the current computation is set to be $P_r=1$.
Initially, the shock wave, followed by a contact discontinuity, moves towards to the right wall. A thin boundary layer is created above the lower wall. The complex shock and boundary layer interaction occurs and results in a lambda-shape shock pattern after the reflecting shock wave from the right wall.
The local mesh and density field at $t=1$ are presented in Fig. \ref{2-viscous-tube-1}. The complex flow structure, including the lambda shock and the vortex configurations, are well resolved by the current compact GKS with a mesh size $h=1/400$. A quantitative verification for the result is also given in Fig. \ref{2-viscous-tube-2}. The density distribution along the lower wall is presented. The result from non-compact high-order GKS \cite{li2010-HGKS} with $h=1/720$ structured mesh is used as the reference solution. The result of third-order CPR-GKS scheme \cite{CPR-GKS-3rd} with $h=1/500$ triangular mesh is also plotted. The current 4th-order compact GKS has a better resolution even with a coarse mesh.
The third-order CPR-GKS needs trouble cell detection for the shock, which isn't needed in the current GKS.

\begin{figure}[!htb]
	\centering
    \includegraphics[width=0.45\textwidth]{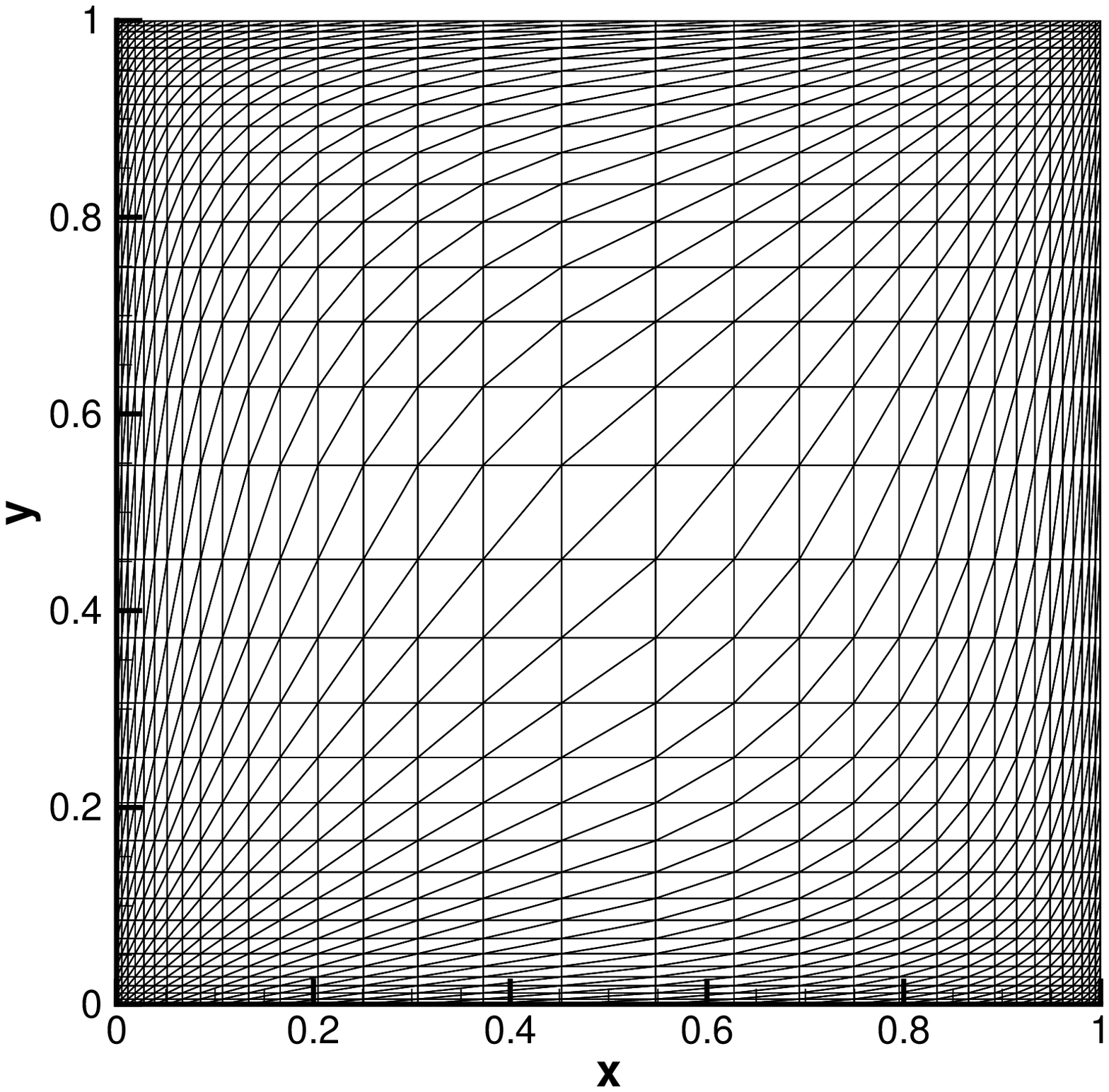}
    \includegraphics[width=0.45\textwidth]{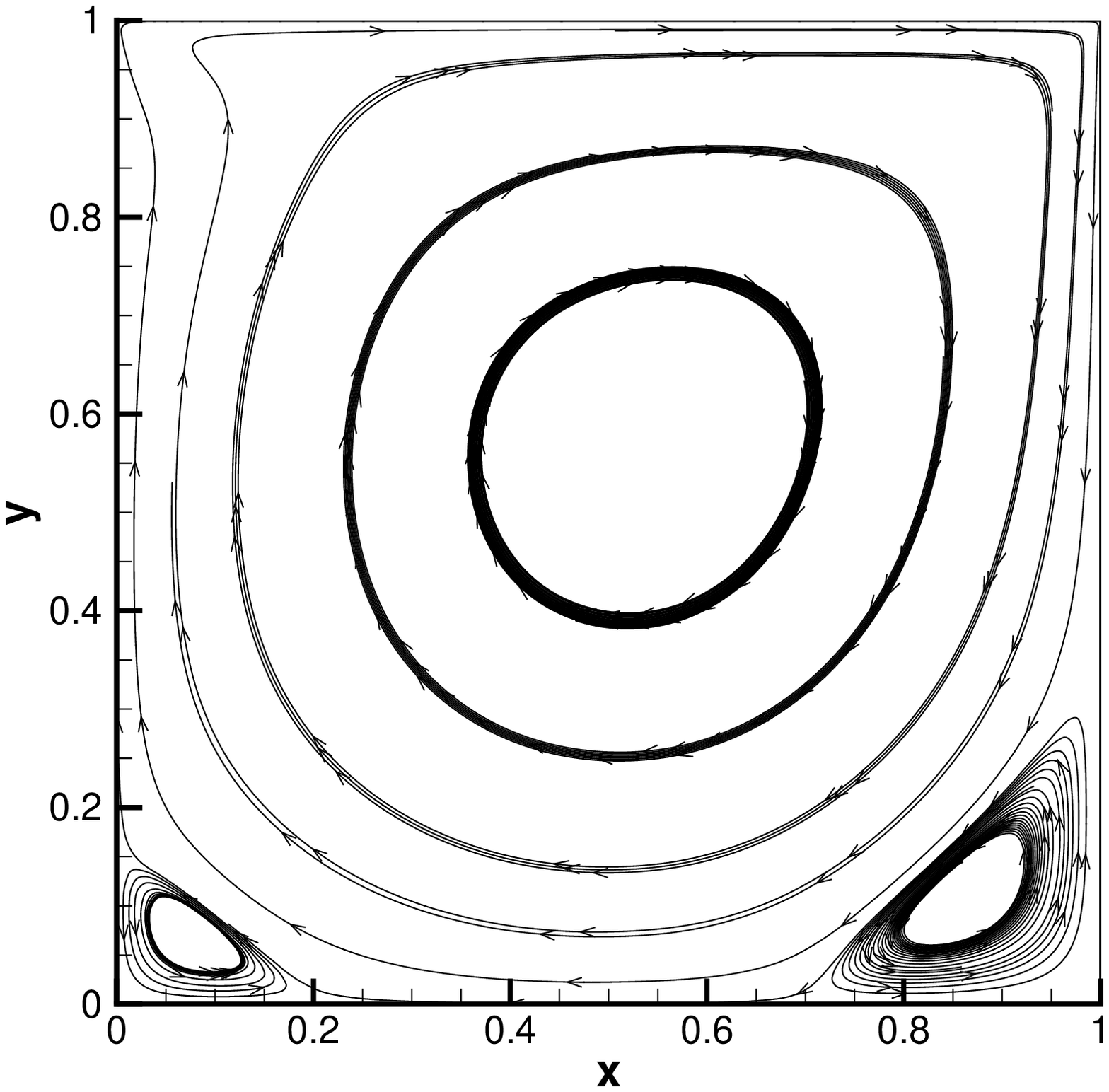}
	\caption{\label{2-cavity-1} Cavity flow: the computational mesh and the streamlines.}
\end{figure}

\begin{figure}[!htb]
	\centering
    \includegraphics[width=0.45\textwidth]{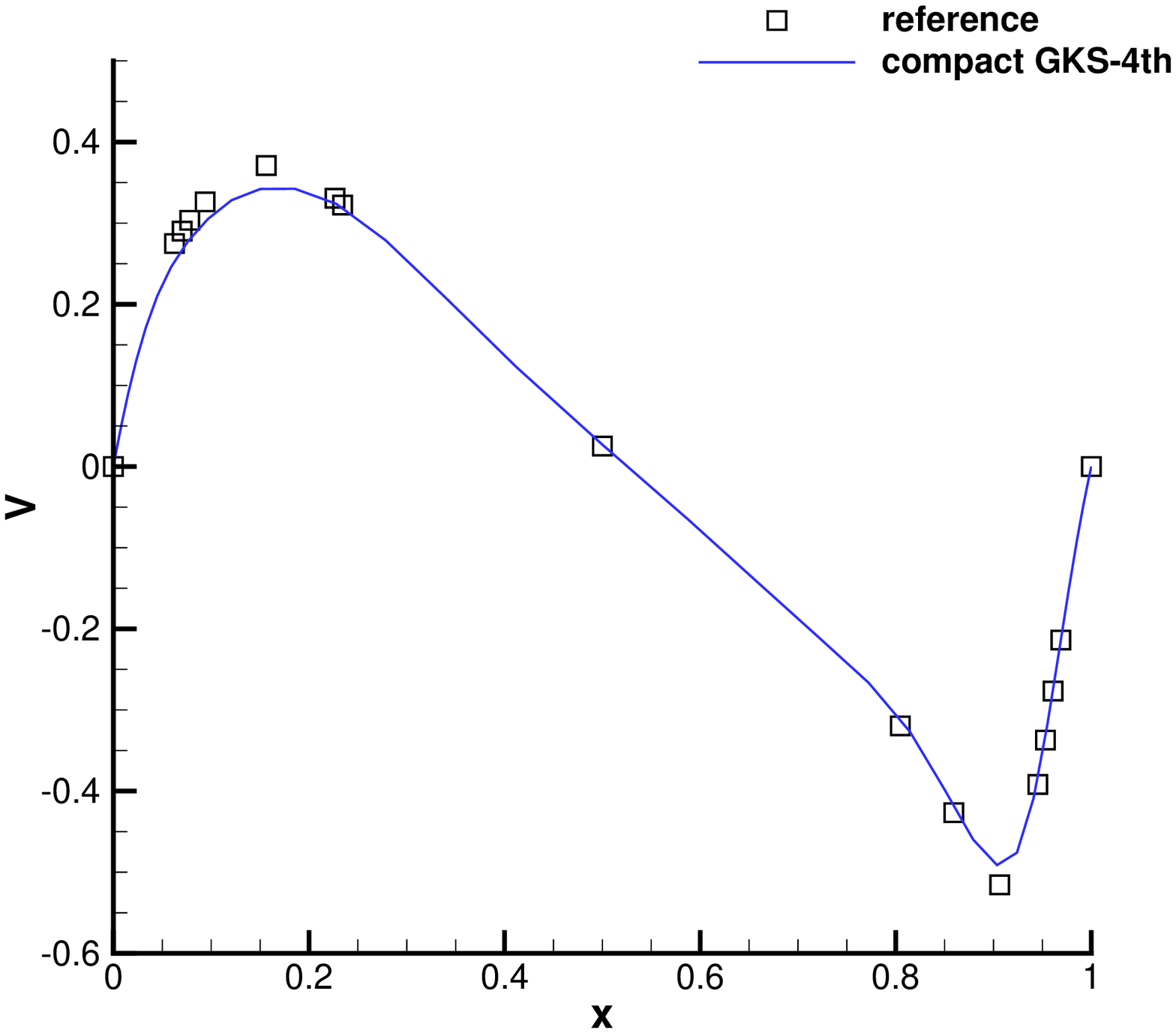}
    \includegraphics[width=0.45\textwidth]{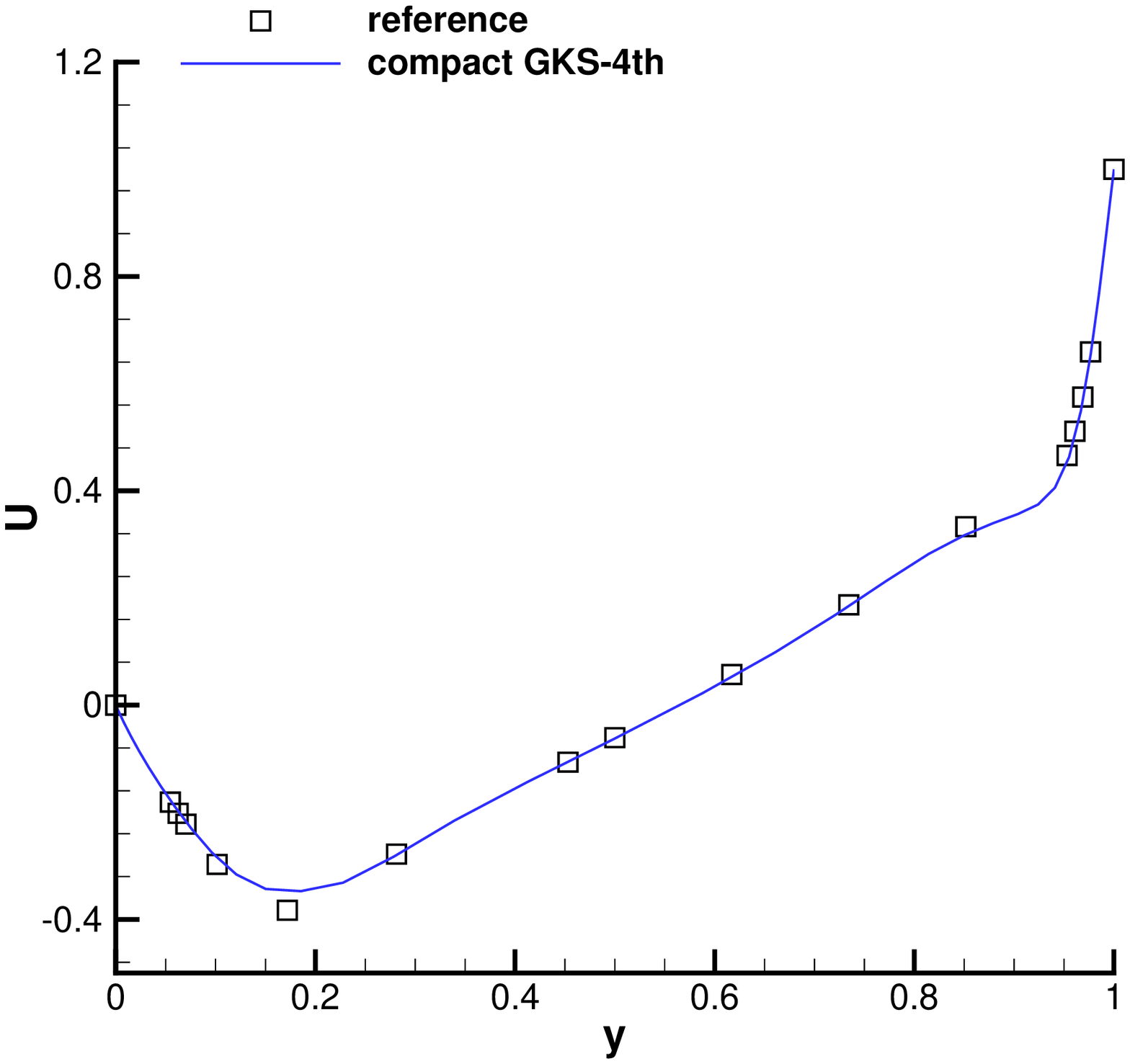}
	\caption{\label{2-cavity-2} Cavity flow: V-velocity (left) distribution along $y=0.5$ and U-velocity (right) distribution along $x=0.5$ by
$33 \times 33 \times 2$ triangular mesh points.}
\end{figure}

\subsection{Lid-driven cavity flow}
The lid-driven cavity problem is a test case for incompressible viscous flow. The fluid is bounded in a unit square where the top boundary is  moving with a uniform speed $U_0=1$ and temperature $T_0=1$.
The corresponding Mach number is $Ma=U_0/\sqrt{\gamma R T_0}=0.15$. The non-slip and isothermal boundary conditions are imposed on all boundaries with the wall temperature $T_w=T_0$. The initial flow is stationary with density $\rho_1=\rho_0$ and temperature $T_1=T_0$. The computational domain is $[0,1]\times[0,1]$. The case of $Re=1000$ is tested.
The computational mesh and the streamlines are shown in Fig. \ref{2-cavity-1}. A total of $33\times33\times 2$ mesh cells are used, and the stretching rate in mesh size is $1.2$. The mesh is refined close the wall and the minimum mesh size is about $h=0.0052$.
The velocities distribution along horizontal and vertical center lines are shown in Fig. \ref{2-cavity-2}. The numerical results agree well with the reference solutions.

\section{Conclusion}

In this paper, based on the framework of reconstruction, evolution, and projection procedures in a numerical scheme,
we present a direct modeling for the construction of high-order compact scheme.
The current difficulty in the compact scheme development is mainly coming from the
requirement of reliable flow variables updates, such as the averaged conservative flow variables and their gradients,
and the corresponding high evolution model in both space and time, where the low order dynamic model of Riemann solver
cannot accomplish the mission.
It becomes more problematic once a unsteady discontinuous shock moves across a cell interface within a time step.
In order to evolve reliable flow variables inside each control volume, two recipes through the direct modeling have been proposed.
One is to evolve the discontinuous flow variables at a cell interface for the update of cell averaged gradients.
The other is to limit time accurate flux function once it is a discontinuous function of time as
a shock passing through the cell interface within a time step.
Similar to the nonlinear limiter in space, such as the WENO scheme in reconstruction, for the first time the limiting process is introduced in the
high-order time derivatives of the flux function as well.
In other words, the S2O4 discretization is generalized to the discontinuous case.

In summary, the direct modeling of CFD is based on the evolution solutions in Eq.(\ref{conservation}) and (\ref{slope}).
In order to close the equations, the development of high-order evolution model becomes necessary.
In order to construct reliable evolution model, the nonlinearly limited flux function in time and accurately evolved multiple cell interface values in space have been proposed in this paper.
Equipped with the above two recipes, the developed high-order compact GKS shows a significant favorable performance
in comparison with previous compact GKS without them, in terms of robustness and efficiency of the schemes.
For example, the computation of a Mach number $20$ flow impinging on a cylinder can be started from time $t=0$ without any difficulty in
the current scheme for both structured and unstructured mesh.
At the same time, a large CFL number can be used in the scheme, such as the CFL number $0.8$ in most numerical examples.
Based on the direct modeling, this paper provides a general framework in the construction of high-order compact schemes.
The essential point is to follow the discretized conservation laws and to construct high-order evolution model to close the discretized
governing equations.
This kind of direct modeling has been successfully used in the construction of multiscale method as well \cite{xu2021}.

\section*{Acknowledgements}

The current research is supported by National Numerical Windtunnel project, National Science Foundation of China (11772281, 91852114), and
Department of Science and Technology of Guangdong Province (Grant No.2020B1212030001).


\bibliographystyle{ieeetr}
\bibliography{AIAbib}

\end{document}